\documentclass[12pt]{iopart}
\usepackage{graphicx,subfigure}
\usepackage{bm}        
\usepackage[mathscr]{euscript}
\begin{document}
\newcommand{\braket}[3]{\bra{#1}\;#2\;\ket{#3}}
\newcommand{\projop}[2]{ \ket{#1}\bra{#2}}
\newcommand{\ket}[1]{ |\;#1\;\rangle}
\newcommand{\bra}[1]{ \langle\;#1\;|}
\newcommand{\iprod}[2]{\bra{#1}\ket{#2}}
\newcommand{\logt}[1]{\log_2\left(#1\right)}
\def\cI{\mathcal{I}}
\def\cC{\tilde{C}}
\newcommand{\cx}[1]{\tilde{#1}}

\def\be{\begin{equation}}
\def\ee{\end{equation}}
\title[Interference of the signal of a local process with the quantum state propagation]{{Interference of the signal from a local dynamical process with the quantum state propagation in spin chains}} 
\author{Saikat Sur$^1$ and V. Subrahmanyam$^2$}
 \address{ Department of Physics, Indian Institute Of Technology,  Kanpur-208016, India}
 
 \ead{saikatsu@iitk.ac.in$^1$ and vmani@iitk.ac.in$^2$}
\date{\today}
\begin{abstract}
The effect of a local instantaneous quantum dynamical process (QDP), either unitary or non-unitary, on the quantum state transfer through a unitary Hamiltonian evolution is investigated for both integrable and non-integrable dynamics. There are interference effects of the quantum state propagation and the QDP signal propagation. The state transfer fidelity is small for further sites, from the site where the information is coded, indicating a finite speed for the propagation of the quantum correlation. There is a small change in the state transfer fidelity for the case of non-unitary QDP intervening the background unitary dynamics.  In the case of unitary QDP, the change is more pronounced, with a substantial increase in the fidelity for appropriate sites and times. For the non-integrable case, viz. a kicked Harper model,
the state transfer fidelity is quite large for further sites for short times, indicating a finite speed for the propagation of the
quantum correlation cannot be defined.
 
\vskip 0.5cm
\hfill {PACS: 03.65.Ud 03.67.Bg 03.67.Hk 75.10.Pq}
\end{abstract}
\maketitle

 \maketitle
\section{ Introduction}

Quantum spin chains have been investigated, over the last few years, from the viewpoint of quantum information and communication. A quantum spin chain serves as a possible
channel for the quantum state transfer\cite{sbose,vs1,christandl,fazal}. These systems have been studied extensively, as exactly-solvable condensed matter physics systems, for  studying spin ordering of novel spin states, and quantum critical behaviour.\cite{sachdev, jordan, bethe}. These studies deal with the quantum dynamics of initial many-body states through the Schroedinger time evolution\cite{chandra},  and the statistical mechanics of the phase transitions\cite{sachdev,chandra}.

The spin chain dynamics can be treated viewing the spin chain as a closed system, with a given initial state with a distribution of correlations. Through the time evolution the quantum correlations can be redistributed and the spin chain can exhibit a variety of multi-party quantum correlation and entanglement structures.
The wide spectrum of  the dynamics investigation of spin chains includes the study of magnon bound states and scattering\cite{ganahl,fukuhara}, spin current and relativistic density wave
dynamics\cite{steinigeweg,foster}, quantum correlations after a quench\cite{cazal} and light-cone in entanglement spreading\cite{manmana}. The effect of unitary evolution of the quantum spin chain can be viewed as, in the quantum information language, various global and local multi-qubit gate operations acting on the many-qubit system. The system will undergo a
redistribution of entanglement and quantum correlations\cite{levitin,burrel,cianciaruso,sarul,cheneau} through the unitary evolution. However,  a multi-party qubit system in general can experience non unitary or incoherent processes \cite{nc} acting on some qubits locally or globally, that can lead to decoherence. Investigating the effect of local decohering process on the quantum correlations and the entanglement structure is difficult in general, as traditional approximation tools like the time-dependent perturbation theory are not applicable for these systems.  In a simple example of local decohering process intervening the Schroedinger evolution of quantum states is investigated\cite{ssvs}, where a  the Hamiltonian unitary evolution of a many-qubit quantum state is interrupted by a local instantaneous quantum dynamical process (QDP) on a certain qubit at a certain epoch of time. The QDP can be a non-unitary decohering operation or a unitary operation different from the back ground evolution.  The QDP signal, the effect of the occurrence of the process at a given time and location,  and its propagation have been investigated using both magnetisation conserving and non-conserving dynamics for  initial states with and without entanglement for various model Hamiltonian dynamics \cite{ssvs}. The signal propagation speed in general depends on the type of interaction of the qubits, the strength of the interaction, the spatial range of the interaction and the initial state.

The quantum state transfer protocol\cite{sbose, vs1,christandl,mhy} studied for the spin chains relies on the Hamiltonian evolution for a faithful communication and detection of the state from an initial qubit location to a target qubit location. The average fidelity, averaged over all possible input qubit state, of the state transfer, is a figure of merit of the protocol. The fidelity depends on the interaction parameters of the Hamiltonian, and in general it is oscillatory as a function of the time of evolution as a result of quantum interference. In general, the decoherence of individual spins,  in the quantum spin channel during  the state transfer,  reduces the fidelity; this has been discussed in \cite{cai}. Protected quantum state transfer through a noisy channel by coupling the end registers has been proposed in \cite{qin}. The state transfer fidelity can have non trivial effects due to the QDP intervention of the unitary evolution, and the fidelity can increase or decrease depending on the location, time and parameters  of the QDP; the detailed investigation of these effects is the main focus of this paper. In Harper model, the transverse potential is a periodic kick, that causes every qubit to undergo a  coherent operation,  during the evolution with the  background XY dynamics. Depending on the strength and the kicking period,  the dynamics is regular or chaotic\cite{arul3,lima}. Thus, we cam discuss the state transfer in a non integrable model, and  contrast it with the behaviour for the integrable model. 

This paper is arranged as follows. In Section 2, we outline the quantum state transfer protocol and the computation of the transfer fidelity. We will study the effect of a local decohering process in Section 3, followed by a local coherent process in Section 4, the computation of the fidelity using one and two qubit Green's functions.  We will discuss a local decohering process in non-integrable systems in Section 5.

\section{{Quantum State transfer through Unitary Dynamics }}

Faithful and faultless transmission of quantum states between two distant locations is the main challenge for quantum communication and information processing protocols. The quantum spin chains as channels of quantum state transfer was proposed and investigated\cite{sbose, vs1, mhy}. The spin of the electrons are basic examples of qubits, viz. two-level quantum systems, and the spin chains can be realised physically in magnetic systems. One-dimensional spin models have been studied extensively using the Heisenberg-XY or transverse-field Ising model Hamiltonians, due to their exact solvability
using the Bethe ansatz technique or Jordan-Wigner fermion technique respectively. In this paper, we will be using the Heisenberg dynamics for the most part, and will study a variant of
transverse-XY model in the last section. 

In a standard state transfer protocol, a direct-product many-qubit initial state is prepared, at time $t=0$, with the first spin  in a desired state (or the information is coded into the state of the first spin). The many-qubit state is evolved through time using the Hamiltonian Schroedinger dynamics
to a final state at time $t$. Then the desired state is to be recovered from a certain target spin, the probability of the state transfer is measured by the fidelity, normalised to unity, computed by taking the overlap of the target state with the desired state. A large value of the state transfer fidelity for a spin at a particular location at a particular time implies the efficacy of the 
state transfer protocol. In this paper, we will deviate from the standard protocol, and intervene the smooth Heisenberg-XY dynamics by a local quantum dynamical process (QDP) at a
given location and time, viz. an instantaneous quantum operation is carried out on the many-qubit state at a time $t=t_0$, between the coding time $t=0$ of the quantum information at the first spin, and the recovery or readout time $t$ of the quantum information from a different spin.
In this paper the focus would be on the effect of QDP on quantum state transfer fidelities. We will see that local QDPs can decrease or increase the fidelity depending upon their locations and time. Also, we will discuss the state transfer and  the effect of local QDP with a non integrable background Hamiltonian dynamics, and investigate the signature of quantum chaos in Sec. 5.

Let us consider a one-dimensional array of N qubits interacting through a nearest-neighbour exchange interaction, and a particular qubit state has to be transferred from one end to the other. The protocol encodes the information to be transferred in to the state $|\phi\rangle = \alpha| 0\rangle +\beta |1\rangle$. The first qubit  is initialised to this state at one end, the many qubit initial state is $|\psi(0)\rangle = |\phi 00...0\rangle$,  a direct product of the state $| \phi ~\rangle$ for the first qubit and the state $|0\rangle$ for all other qubit. The unitary time evolution will transform the many qubit state to $|\psi(t)\rangle$ using the Hamiltonian dynamics. The desired qubit $|\phi\rangle$  can spread out to other qubits during the evolution, 
i.e., the desired qubit state can travel to the other locations with a some speed and probability.  The speed associated with such transfer is inversely related to the interaction strength  between the sites. For an efficient state transfer, the target state  $\rho_l$, is the reduced density matrix of the $l$'th qubit, calculated by taking a trace over all other qubits from the
many qubit state, $\rho_l = tr^\prime |\psi(t\rangle \langle \psi(t)|$, should be the same as the desired state $\rho_{\phi}=|\phi\rangle\langle \phi|$. The
 state transfer fidelity gives a probabilistic measure of the state transferred from one site to another and is quantified by taking overlap of the target and the desired states, given by
\begin{equation}
\mathscr{F}_{l,\phi} \equiv Tr \rho_{\phi} \rho_l= \langle \phi | \rho_l| \phi\rangle. 
 \end{equation}
Now, the fidelity shown above depends on the initial state $\phi$ , apart from the target spin location and time, and the interaction parameters of the Hamiltonian dynamics. A figure of
merit measure for the efficacy of the protocol is to average the above over all possible initial qubit states, i.e., over the surface of the Bloch sphere representing all qubit pure states.

 The first exactly-solvable and integrable non trivial models of interacting quantum spins is a one-dimensional chain of spins interacting with their nearest neighbour Heisenberg exchange interaction, known as the Heisenberg model.  We will use the Pauli operator $ \vec \sigma_l $ to represent the different components of the spin operator $\vec S_i$ at $i$'th site. Let us consider a one-dimensional chain of $N$ spins interacting through the nearest-neighbour anisotropic Heisenberg model. The Hamiltonian is given by,
 \begin{equation}
H = -J\sum_{i}({\sigma}^x_{i}{\sigma}^x_{i+1}+{\sigma}^y_{i}{\sigma}^y_{i+1}+\Delta {\sigma}^z_{i}{\sigma}^z_{i+1}).
 \end{equation}
where $J$ is the exchange interaction strength for the nearest-neighbour spins, and $\Delta$ is the anisotropy strength. As all the three Pauli spin matrices appear in the Hamiltonian, an exchange interaction of neighbouring spin is implied in all three spin dimensions.  The model exhibits ferromagnetic (antiferromagnetic) behaviour in the ground state for $\Delta >0 (\Delta <0)$. The ground state and all the excited states are known, and can be found using the Bethe ansatz\cite{bethe}.  Let us use the basis states for the $i$'th spin as $|0\rangle$ (up-spin state) and  $|1\rangle$(down-spin) states, denoting the eigenstates $\sigma_i^z$ with eigenvalues +1 and -1 respectively. The basis states for the many-qubit system can be chosen to be the direct products of the basis states of each spin. The z-component of the total spin $\Sigma \sigma_i^z$ is a constant of
motion,  which implies that the eigenstates will have a definite number of down spins.  The many-qubit basis states with $l$ down spins can also be labeled by the locations ($x_1,x_2..x_l$) of the $l$
down spins, where the set is an ordered set with $x_1<x_2$ and so on. An eigenstate with $l$ down spins, a $l-$magnon state,  can be written as a superposition of the basis states as,
\begin{equation}
|\psi\rangle = \sum_{x_1,x_2..x_l}\psi ({x_1,x_2..x_l}) |x_1,x_2..x_l\rangle
\end{equation}
where the eigenfunction $\psi ({x_1,x_2..x_l})$ denotes the wave function amplitude for the corresponding basis state. The eigenfunction is given by the Bethe Ansatz\cite{izyu},
labeled by the the set of momenta
$(p_1,p_2..p_l)$ of the down spins, which are determined by solving algebraic Bethe ansatz equations, with periodic boundary conditions. 
There is only one zero-magnon state $|F\rangle =|00..0\rangle$, which is just a ferromagnetic ground state with all the spins polarized along one direction. It is straightforward to see that it an eigenstate of the above Hamiltonian with energy $\epsilon_0=-N J$ for periodic boundary conditions (closed chain), and $\epsilon_0=-J(N-1)$ for open boundary conditions (open chain). Starting from $|F\rangle$, one-magnon excitations can be created by turning any one of the spins, giving $N$ localised one-magnon states, which can be labeled by the location of the down spin. One-magnon eigenstates are labeled by the momentum of the down spin, the eigenfunction is given by,
\begin{eqnarray}
\psi_p^x = & \sqrt{\frac{1}{N}} e^{i px}; p ={ 2\pi I\over N}, {\rm for ~a~ closed ~chain} \nonumber \\
\psi_p^x = &\sqrt{\frac{2}{N+1}} \sin (px); p= {\pi I\over N+1},  {\rm for ~an ~open~ chain},
\end{eqnarray}
where the momentum $p$ is determined by an integer $I=1,2,..N$ for both cases. The one-magnon eigenvalue is given by $\epsilon_1(p)=\epsilon_0-2J\cos{p}$.
The interaction strength $J$ determines the hopping of the down spins to neighbouring sites, and the interaction of the two down spins is determined by $\Delta$. The one-magnon eigen energies are independent of $\Delta$ as the states carry
only one down spin. The two-magnon (l=2) and other eigenstates (l $>$ 2) include both scattering states and bound states of the down spins. The eigenfunction for the two-magnon eigenstate
is labeled by two momenta $p_1$ and $p_2$, is given by 
\begin{equation}
\psi_{p_1,p_2}^{x_1,x_2}= A(p_1,p_2)(e^{ip_1x_1}e^{ip_2x_2} + e^{i \theta (p_1,p_2)}e^{ ip_1 x_2}e^{ip_2 x_1}),
\end{equation}
where $A(p_1,p_2)$ is a normalisation factor.
The two terms in the wave function differ by a permutation of the positions $x_1$ and $x_2$. The phase factor $\theta (p_1,p_2)$ depends on the moment and the interaction strengths, and is given by
\begin{equation}
\tan{\theta (p_1,p_2) \over 2}= { \frac{\Delta \sin[(p_1-p_2)/2]} {\cos[(p_1+p_2)/2] -\Delta \cos[(p_1-p_2)/2] } }.
\end{equation}
 The two momenta are determined by the Bethe ansatz equations,
\begin{equation}
p_1N = 2\pi I_1 + \theta(p_1,p_2), ~ p_2N = 2\pi I_2 - \theta(p_1,p_2),
\end{equation}
where $I_1,I_2$ are any integers. The two-magnon eigenvalues are given by $\epsilon_2(p_1,p_2)=\epsilon_0+2(\Delta-\cos{p_1})+2(\Delta - \cos{p_2})$. The two-magnon scattering states have been studied for the spin-independent scattering\cite{banchi}. The detail of solving the
above Bethe ansatz equations for the two-magnon scattering and bound states is discussed in the Appendix A, along with the computation of the time-dependent Green's functions. 
\begin{figure}
\center{
 \includegraphics[width=0.75\textwidth]{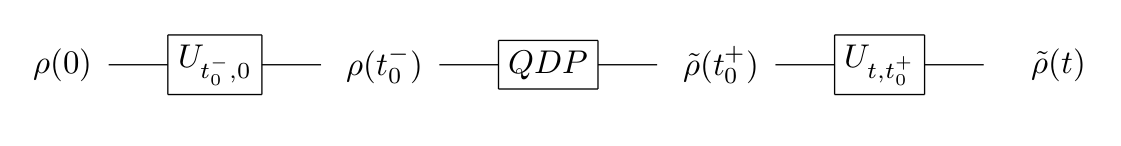} }
 
\caption{ \small 
The evolution of the initial state $\rho(0)$ in three steps: The state evolves to $\rho(t_0^-)$ through a unitary process using the Heisenberg Hamiltonian till $t=t_0^-$ , the state become $\tilde \rho(t_0^+)$ through an instantaneous local QDP at $t=t_0$, and finally it evolves to $\tilde \rho(t)$ through the same unitary process.}

\end{figure}

  We will see no dynamics effects if the initial state is this eigenstate of the Hamiltonian. 
  To study
  the dynamics of an arbitrary initial state, we will consider a linear combination of zero-magnon and one-magnon  states that are not eigenstates of the Hamiltonian, in conjunction with the QDP dynamics below. 
Let us consider a spin chain with open boundary condition and the initial state given by,
 \begin{equation}
|\psi(0)\rangle =  \alpha|00...0\rangle+\beta|100..0\rangle = \alpha|F\rangle+\beta|x = 1\rangle,
 \end{equation}
 where $x$ denotes location of the down spin. The above state is a linear superposition of an eigenstate of the Hamiltonian with eigenvalue $\epsilon_0$ and a one magnon state which can be written as a linear combination of one-magnon momentum eigenstates $|p\rangle$ of the Hamiltonian  with eigenvalues $\epsilon_1(p)$, we have
  \begin{equation}
|\psi(0)\rangle = \alpha|F\rangle +\beta \sum_{p}\psi_p(1)|p\rangle,
 \end{equation}
 Now, the time evolution of the system will transport the single down spin from the first site  to other sites and the  state after a time $t$ becomes,
   \begin{equation}
|\psi(t)\rangle =  \alpha e^{-i\epsilon_0 t}|F\rangle +\beta \sum_{y}G^{y}_1(t)|y\rangle,
 \end{equation}
  where the time-dependent function $G^{x'}_{x}(t)$  \cite{ssvs} is given in terms of the wave functions defined above as,
   \begin{equation} 
 G^{x'}_{x}(t)=\sum_{p}\psi_{p}^x  \psi_p^{x'*} e^{-it\epsilon_1 (p)}.
  \end{equation}
  Using the one-magnon eigenfunctions, shown in Eq.4, we can express the Green's function as,
  \begin{equation} 
G^{x'}_x(t) = \frac{N}{2 \pi}\int_{0}^{\pi} \psi^x_p \psi^{x'*}_{p} e^{-it\epsilon(p)} dp= (-i)^{(x-x')}J_{x-x'}(2t)-(-i)^{x+x'}J_{x+x'}(2t),
  \end{equation}  
 for an open chain, and we have
  \begin{equation} 
 G^{x'}_x(t)  = (-i)^{x-x'}J_{x-x'}(2t) ,
  \end{equation}  
  for a closed chain.
$J_x(y)$ is the Bessel function of integer order $x$ and argument $y$. The reduced density matrix (RDM)of the $l^{th}$ qubit is defined by taking trace of the whole quantum  state over all qubits except the  $l^{th}$ one,  as
$\rho_l = Tr^{\prime }\rho (t)$, where $\rho (t)=|\psi(t)\rangle\langle \psi(t)|$, is the time-evolved many-qubit state. Using the basis $|0\rangle, |1\rangle$ for
the $l$'th qubit, the reduced density matrix is given by,
\begin{equation}
\rho_l(t)  = \left ( \begin{array}{cc}
     1-x_l(t) &  y_l(t) \\
   y^{*}_l(t) & x_l(t)  \\
\end{array}
\right),
\end{equation}
where the elements of the RDM  are  time dependent functions that can be calculated from Eq.10 as,  
  \begin{eqnarray} 
 x_l(t) = \langle 1|\rho_l(t)|1\rangle = |\beta|^2|G^l_{1}(t)|^2,\nonumber\\ y_l(t) = \langle 0|\rho_l(t)|1\rangle =\alpha\beta^*e^{-i\epsilon_0t}G^{*l}_1(t).
  \end{eqnarray}
 The state transfer fidelity for the $l^{th}$  site as a function of time is given by,
   \begin{equation} 
\mathscr{F}_{l,\phi} (t) =|\alpha|^2(1-x_l(t))+|\beta|^2 x_l(t)+ 2 Re (\alpha^{*}\beta y_l(t)).
  \end{equation}
  
The quantity $\mathscr{F}_{l,\phi} (t)$ depends on the parameters $\alpha$ and $\beta$ of the initial state. 
  The averages over the parameters in the above equations are given as  $\overline{|\alpha|^2} =  \overline{|\beta|^2} = \frac{1}{2}$, $\overline{|\alpha|^2 |\beta|^2} = \frac{1}{6}$and  $\overline{|\beta|^4} = \frac{1}{3} $. Taking the average over all possible pure states on the Bloch sphere characterised by  $\alpha$ and $\beta$ the average state transfer fidelity\cite{sbose} becomes,
\begin{equation} 
\mathscr{F}_l(t)= \frac{1}{2}+\frac{1}{6}|G^l_{1}(t)|^2+\frac{1}{3}Re(e^{i\epsilon_0t}G^l_{1}(t)).
  \end{equation}
The state transfer fidelity $\mathscr{F}_l(t)$ with unitary dynamics is shown in Fig. 2(a) as a function of time for different site indices. Since the function $G^l_1(t)$ is a Bessel function with time as argument it falls inversely with time and  the fidelity of state transfer decreases with time for a particular site and over a large time saturates to the value 0.5. It also deceases with the increase in the distance between the first site and the decoding site l. The maximum fidelity for the site $l$ is obtained around time $l/2$ (here, we have taken interaction strength $J= 1/2$). So, the speed of state transfer is equal to the value of interaction strength $J$ between the spins.

 \begin{figure*}[t!]
    \begin{center}
        \subfigure[]{%
            \label{fig:first}
            \includegraphics[width=0.36\textwidth]{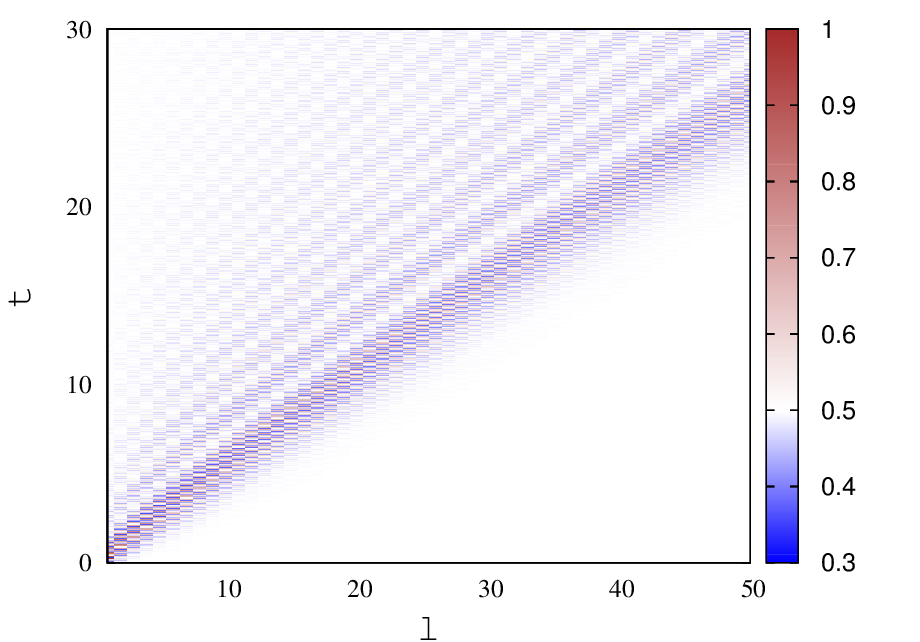}
        }%
        \subfigure[]{%
           \label{fig:second}
           \includegraphics[width=0.36\textwidth]{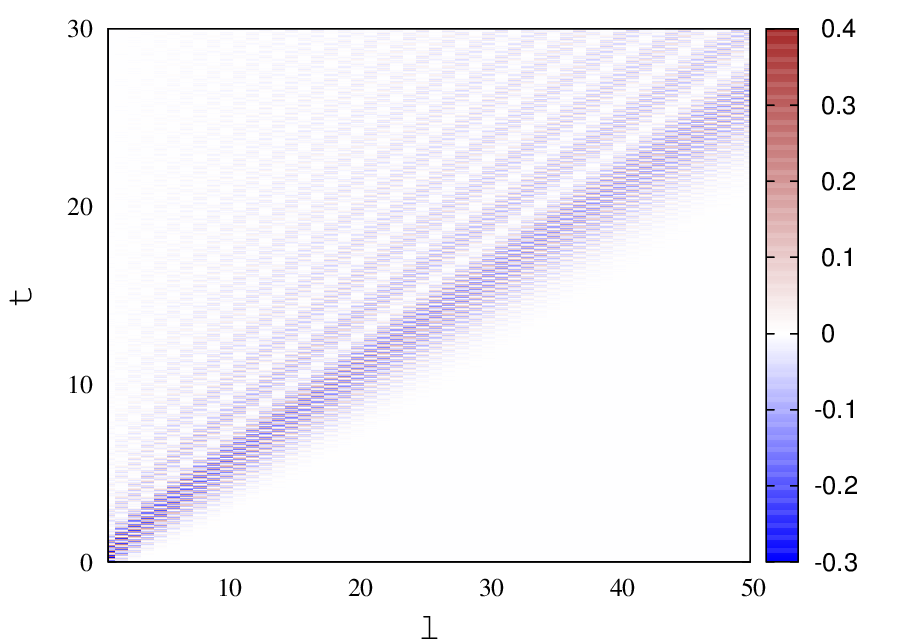}
        }%
        \subfigure[]{%
          \label{fig:third}
            \includegraphics[width=0.36\textwidth]{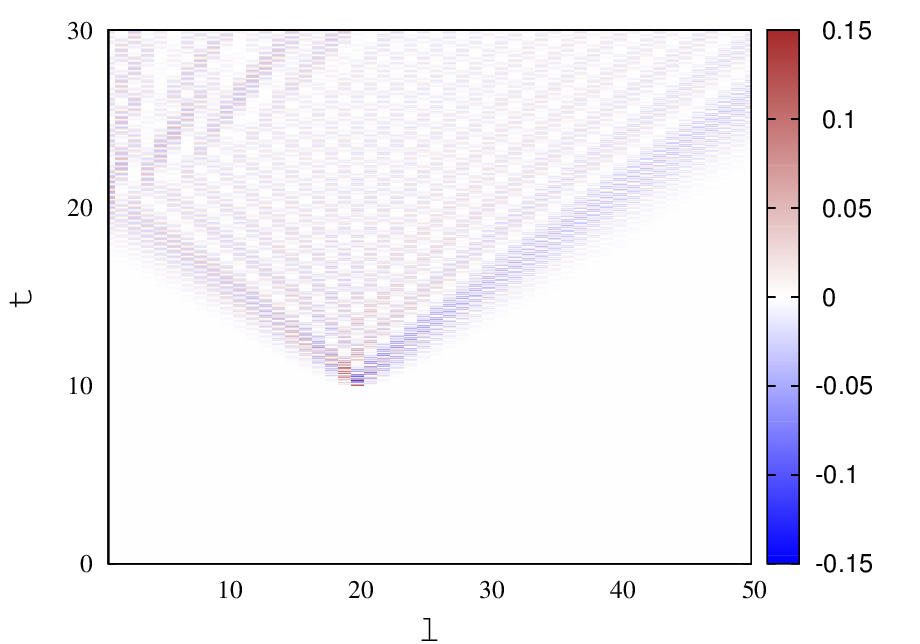}
        }\\%
        \subfigure[]{%
            \label{fig:fourth}
            \includegraphics[width=0.36\textwidth]{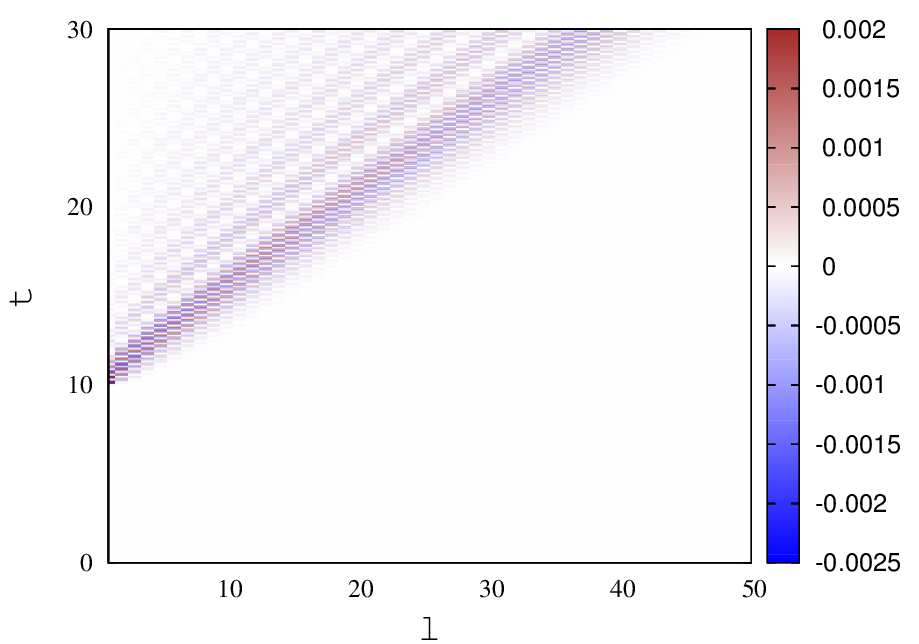}
        }%
        \subfigure[]{%
            \label{fig:fourth}
            \includegraphics[width=0.36\textwidth]{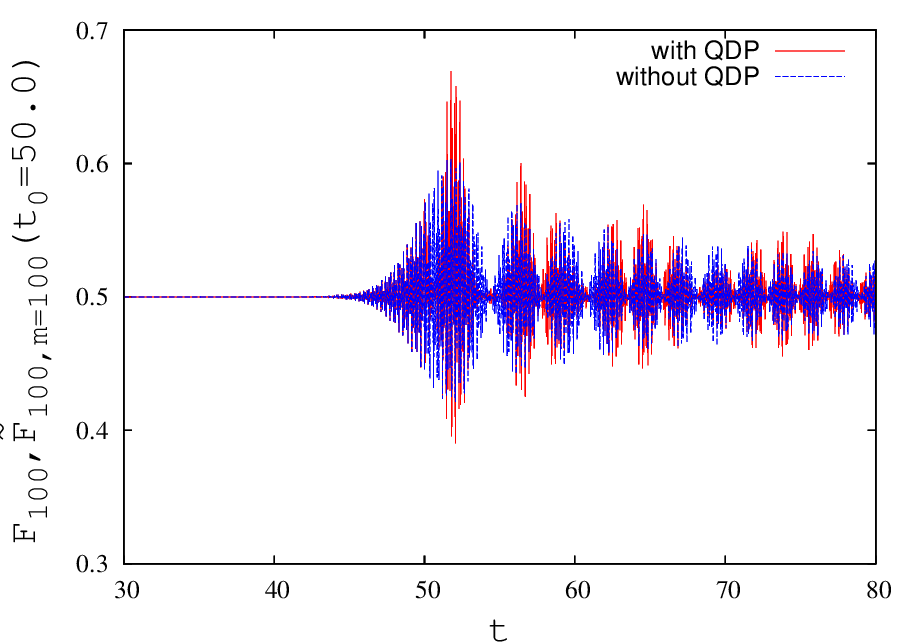}
        }%
        \subfigure[]{%
            \label{fig:fourth}
            \includegraphics[width=0.36\textwidth]{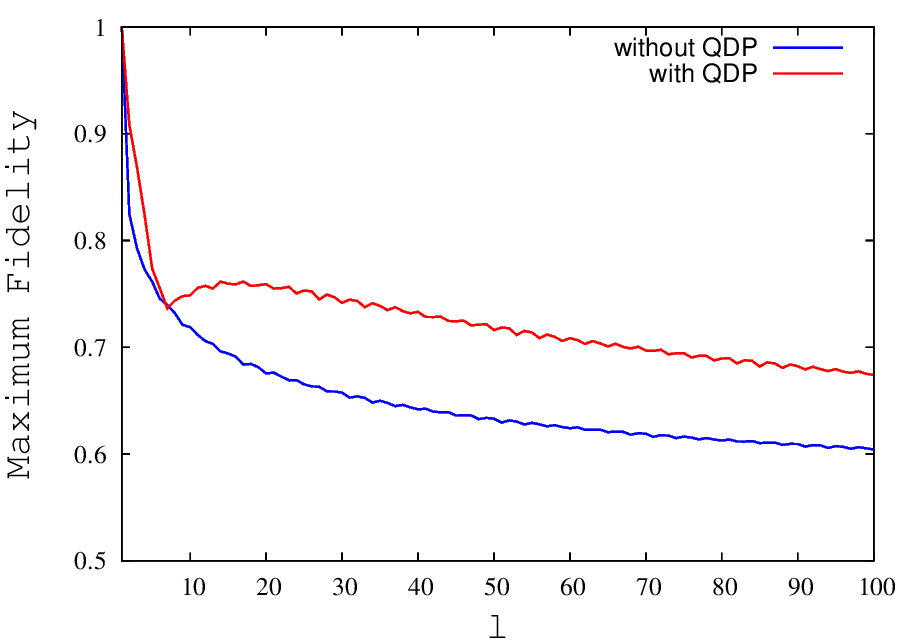}        
        }\\%

 \end{center}
    \caption{\label{fig:fig_2}{\small (a) State transfer fidelity $\mathscr{F}_l(t) $ as function of site index($l$) and time ($t$) for Heisenberg dynamics with with open boundary condition without any decohering process. (b) Differences of state transfer fidelities with and without QDP ($\Delta\mathscr{F}_l,m=1(t; t_0 = 0.0)$) as function of site index($l$) and time ($t$); where the QDP occurs at the site ($m=1$) and time ($t_0 = 0.0 $).(c) $\Delta\mathscr{F}_l, m=20(t; t_0 =10.0)$ as function of site index($l$) and time ($t$); where the QDP occurs at the site ($m=20$) and time ($t_0 = 10.0 $). (d)  $\Delta\mathscr{F}_l, m=1(t; t_0 = 10.0)$ as function of site index($l$) and time ($t$); where the QDP occurs at the site ($m=1$) and time ($t_0 = 10.0 $).(e)Time dependence of the fidelities with QDP $\tilde{\mathscr{F}}_{l,m}(t;t_0)$ and without QDP $\mathscr{F}_l(t)$ for the site $l= 100$, site of QDP $m =100$ and time of QDP $t_0 = 50.0$. (f) Maximum fidelities with and without QDP as a function of site index $l$; the site of QDP $m = l$ and time of QDP $m = l/2$  Number of sites taken $N =100$.}
     }%
   \label{fig:subfigures}
\end{figure*}  

\section{{Local quantum decohering process}}
 We now consider the effect of a local QDP intervening the smooth unitary dynamics, thus effect the state transfer protocol. The time evolution of a general state involves, as we have seen  in the previous section, the coherent movement of the down spins, due to the unitary dynamics generated by the Heisenberg Hamiltonian we are considering.  A local incoherent process or QDP can be any single qubit quantum  gate operation,  for example a local projective measurement\cite{nc,peres}. The QDP we consider  below occurs instantaneously at a time $t=t_0$, operating on a certain spin, thus changing the whole many-qubit state instantaneously. That is is the QDP occurs between the time  $t=t_0^-$ and the time $t=t_0^+$. The dynamics is unitary for $t<t_0^-$, and again for $t>t_0^+$, and it is given by the Heisenberg Hamiltonian. The QDP can be non-unitary, which causes explicit decoherence of the many-qubit state, considered in this section, or the QDP can be another coherent quantum operation that will be discussed in the next section. First, we consider a non-unitary QDP that is a local decohering process, as an example,  a projective measurement of $\sigma_m^z$ which has two outcomes, corresponding projectors $P_0=(1+\sigma_m^z)/2$ and $P_1=(1-\sigma_m^z)/2$.  A more general measurement operators, for example measuring an arbitrary component of $\vec \sigma_m$, have been considered and seen to be qualitatively similar\cite{ssvs}. Thus, now we have marked three qubits, at three different locations, namely, the first qubit where the initial information is coded, the $l$'th qubit where we recover the desired state transfer, and the $m$'th qubit where a local QDP occurs. We will see below the  case of $l>m$, in which both the desired state from the first spin and the effect of QDP will have to travel in the same direction to reach the recovery site, and the case of $m>l$ in which the two effects travel to the recovery site from opposite directions. There will be interesting quantum interferences effects, particularly if the QDP is a coherent operation that will be discussed in the next section.

The initial state $\rho(0)=|\psi(0) \rangle\langle \psi(0)|$ undergoes three different evolutions here in three steps, as depicted in Fig.1 In the first step, the state is evolved through the unitary operation using
the Heisenberg Hamiltonian. The evolution of the state up to $t=t_0^-$ results in the state, $\rho(t_0^-)=U_{t_0,0}\rho(0)U_{t_0,0}^\dag$, where the unitary operator is given by $U_{t_2,t_1}=\exp {-i H (t_2-t_1)}$, where we have absorbed the Planck constant in the time itself or set $\hbar=1$. The state is given by,
\begin{equation}
\rho(t_0^-)= e^{-iHt_0}|\psi(0)\rangle\langle \psi(0)|e^{iHt_0}=|\psi(t_0)\rangle\langle \psi(t_0)|.
\end{equation}
We have discussed the computation of $|\psi(t)\rangle$ in the last section, along with the state transfer fidelity.
In the second step, an instantaneous operation of the local QDP at the $l$'th
site, the state transforms from $\rho(t_0^-)$ to
 $\tilde \rho(t_0^+)$. Using the Kraus operator formalism with the two operators $P_0$ and $P_1$, the state is now given by
 \begin{equation}
  \tilde \rho(t_0^+)= P_0 \rho(t_0^-) P_0^\dag + P_1 \rho(t_0^-) P_1^\dag.
  \end{equation}
  As we can see from the above, the state here is a mixed state with two different pure-state components, suffering decoherence due to the occurrence of the QDP at $m$'th site. 
  Just after the QDP, the state can be written in terms of two pure states as,
    \begin{equation}
\tilde{{\rho}}(t_{0^+})=|\tilde \psi_+(t_0)\rangle \langle \tilde \psi_+(t_0)|+|\tilde \psi_-(t_0)\rangle \langle \tilde\psi_-(t_0)|,
    \end{equation}
where, $|\tilde \psi_{\pm}(t_0)\rangle\equiv {1\pm \sigma_m^z\over 2}|\psi(t_0)\rangle$.  
 In the third step, the state undergoes a unitary time evolution from $t=t_0^+$ to time $t$ using the Heisenberg Hamiltonian, we have 
 \begin{equation}
 \tilde \rho(t)= U_{t,t_0} \tilde \rho(t_0^+) U_{t,t_0}^\dag.
 \end{equation}
 In this step, the unitary Hamiltonian evolution occurs for an interval $t-t_0$. Finally, we can rewrite the state using the two pure states $|\tilde \psi^{\pm}\rangle$, as
 \begin{equation}
\tilde{{\rho}}(t)=|\tilde\psi_+(t)\rangle \langle \tilde\psi_+(t)|+|\tilde\psi_-(t)\rangle \langle \tilde\psi_-(t)|,
 \end{equation} 
  where,
    \begin{eqnarray}
|\tilde \psi_+(t)\rangle= \alpha e^{-i\epsilon_0t}|00...0\rangle +\beta \sum_{y'}H^{y'}_{1}(m,t,t_0)|y'\rangle,\nonumber\\
|\tilde\psi_-(t)\rangle=\beta \sum_{y'}K^{y'}_{1}(m,t,t_0)|y'\rangle,
   \end{eqnarray}
   Here, the time-dependent wave functions\cite{ssvs} are given by,
 \begin{eqnarray}
H^{y'}_{y}(m,t,t_0)= \sum_{y''\neq m} G^{y''}_{y}(t_0) G^{y'}_{y''}(t-t_0), \nonumber \\
K^{y'}_{y}(m,t,t_0)=  G^{m}_{y}(t_0) G^{y'}_{m}(t-t_0).
   \end{eqnarray}
 
Now, the elements of  the time dependent RDM  (given in Eq. 14) of the $l^{th}$ site are given by,

 \begin{eqnarray}
\tilde{x}_l(t) = \langle 1|\tilde{\rho}_l(t)|1\rangle =  |\beta|^2(|H^l_{1}(m,t,t_0)|^2+|K^l_{1}(m,t,t_0)|^2), \nonumber \\
\tilde{y}_l(t) =  \langle 0|  \tilde{\rho}_l(t)|1\rangle = \alpha\beta^*e^{-i\epsilon_0 t}H^{*l}_{1}(m,t,t_0).
   \end{eqnarray}

 The average state transfer fidelity averaged over all pure states on the Bloch sphere for $l^{th}$ site is given by,
 \begin{eqnarray} 
\tilde{\mathscr{F}}_{l,m}(t;t_0)= \frac{1}{2}+\frac{1}{6}(|H^l_{1}(m,t,t_0)|^2+|K^l_{1}(m,t,t_0)|^2)+\frac{1}{3}Re(e^{i\epsilon_0t}H^l_{1}(m,t,t_0))\nonumber\\=\frac{1}{2}+\frac{1}{6}|G^l_{1}(t)|^2+\frac{1}{3}Re(e^{i\epsilon_0t}(G^l_{1}(t) -2K^l_{1}(m,t,t_0))-G^{*l}_{1}(t)K^l_{1}(m,t,t_0))\nonumber\\+\frac{5}{6}|K^l_{1}(m,t,t_0)|^2.
  \end{eqnarray}
  The maximum fidelity saturates to the value $1/2$ over a long time $t$ because the function $G^l_1(t)$ is basically a Bessel function which falls as $1/t$  . In case of a local decohering process the dynamics is not very different or in other words the QDP does not "disturb" the system significantly. 
  Hence, the difference in fidelity with and without QDP for $l^{th}$ site is given by $\Delta\mathscr{F}_{l,m}(t;t_0)  =\tilde{\mathscr{F}}_{l,m}(t;t_0)-\mathscr{F}_l(t) $, we have 
 \begin{eqnarray}
  \Delta\mathscr{F}_{l,m}(t;t_0) = \frac{5}{6}|K^l_{1}(m,t,t_0)|^2 -\frac{1}{3}Re((e^{i\epsilon_0t}+G^{*l}_{1}(t))K^l_{1}(m,t,t_0)).
\end{eqnarray}   
   However, the quantum state transfer process (which is a coherent or unitary process) and the effect of the QDP (which is a decohering process) can interfere with each other and change the fidelity slightly. If the QDP is performed at a particular site at a time when the target state is at site while moving the fidelity changes slightly. From Eq. 27 maximum difference of fidelity with and without QDP can be estimated to be $\frac{5}{6}|K^l_{1}(m,t,t_0)|^2$ at time $t = l/2$, because the second term is merely a oscillating term.\\

   The Fig. 2(b) depicts the change in fidelity for one extreme case where the QDP occurs at the first site just after the evolution starts. In that case the maximum difference in fidelity is around 0.4. However, when the QDP occurs at some other site at a later time this difference is much less. For example in Fig. 2(c) maximum change in fidelity is around 0.15 when the QDP is occurred at the site $m= 20$ at the time $t_0 = 10$. If the QDP occurs at a site where where the basic fidelity without the QDP itself is very small the interference effect is negligible and the fidelity does not change significantly. In Fig.2(d) depicts the case where the QDP occurs at the first site but at much later time $t_0 =10.0$, the difference is seen to be negligible. In Fig. 2(e) the dependence of the fidelity for the site $l = 100$ is plotted with and without the QDP, where the location $m$ of the QDP  is same,  and the time  of QDP  is $t_0$ is $50.0$. Both the fidelities show peaks at a time slightly after $t = 50.0$, but the maximum fidelity with the QDP is slightly greater than that without the QDP.  The time at which the fidelity is maximum for
  is around $t=l/2$, for the QDP occurring at the location $l$.  In Fig. 2(f) the maximum fidelity is shown as a function of site index $l$ obtained with and without QDP,  with the location and the time of QDP being $m = l$ and $t_0 = l/2$. The difference of fidelities is small for sites close to the first site, from where the state transfer has been started. The difference of fidelities is almost constant for further sites, as seen in the figure,  and the difference is around $0.1$.

 \section{{Local quantum coherent process}}
 
 In this section, we will turn our attention to the case of an instantaneous local unitary QDP intervening the unitary dynamics of the Heisenberg Hamiltonian. Similar to the last section, the QDP operates on the $m$'th spin at time $t=t_0$. But unlike the previous section, the unitary QDP does not cause decoherence. The three steps of
 evolution that are involved here, as shown in Fig.1, are all unitary evolutions, as a result the initial pure state evolves into another pure state. The intervening QDP being unitary, it can be generated using a Hamiltonian. The instantaneous local unitary QDP intervening the unitary Hamiltonian evolution can be viewed as an evolution with a kicked Heisenberg Hamiltonian. The new Hamiltonian $\tilde H$ can be written as a sum of two terms, the Heisenberg Hamiltonian $H$, given in Eq. 2 that generates the background unitary evolution, and a magnetic field term $H^\prime= \vec \sigma_m.\hat n$ with a delta-function kick, where $\hat n$ is the direction of the magnetic field at the $m$'th site. We can consider $\hat n$ to be a unit vector in the x-y plane. Thus, the total Hamiltonian covering all the three steps can be written as,
 \begin{equation}
 \tilde H = H + H^\prime \delta (t/t_0-1).
 \end{equation}
 The second term in the Hamiltonian represents the instantaneous QDP operating on the given spin at $t=t_0$.   The unitary evolution operator for $t>t_0$, is a product of three unitary operators corresponding to the three time steps, i.e., $U_{t_0,0}=e^{-iHt_0}$ for the evolution up to $t=t_0^-$, followed by $V_m=e^{-t_0\vec \sigma_m.\hat n}$ for the instantaneous unitary QDP, and $U_{t,t_0}=e^{-i(t-t_0)H}$ for the evolution from $t=t_0^+$ up to time $t$. Thus, an initial state $|\psi(0)\rangle$ prepared at time $t=0$ evolves to the state $|\tilde \psi (t) \rangle= \tilde U_{t,0} |\psi(0)\rangle$,  the evolution operator is given by,
 \begin{equation}
 \tilde U_{t,0}= U_{t,t_0} V_m U_{t_0,0}. \nonumber
 \end{equation} 

 Now,  similar to the non-unitary QDP,  we start with an initial state $|\psi(0)\rangle = \alpha|0..0\rangle + \beta |100.0\rangle$. In the first step, the state
 evolves unitarily upto a time $t_0^-$ with the Hamiltonian $H$,  yielding $|\psi(t_0^-)\rangle$. This part is the same as discussed in the last section,  as shown in Eq. 10. In the second step, the state is changed by an operation of $V_m$,  the $m^{th}$ qubit undergoes a local unitary operation.
 This coherent or unitary process, which is  a local quantum gate operation, is represented by an instantaneous unitary operation that acts on the state of the given qubit between time $t=t_0^-$ and time $t_0^+$. The operation of the unitary operator ${V}_m $ on the basis states of $m$'th spin is given by ,
 \begin{equation} 
{V}_m |0\rangle  = \gamma |0 \rangle + \delta |1\rangle,~
{V}_m |1\rangle = - \delta^* |0 \rangle + \gamma |1\rangle,
 \end{equation}
 where the various amplitudes are related to $t_0$, the components $n_x, n_y$ of the unit vector $\hat n$ as, $\gamma = \cos {t_0 }$ and $(n_y + in_x)\sin{t_0} = \delta$. The state 
 $|\tilde \psi(t_0^+)\rangle$, 
 immediately after the operation at $t=t_0$ is given by, 
 \begin{eqnarray} 
|\tilde\psi (t_{0^+})\rangle = {V}_m [\alpha e^{-i\epsilon_0 t_0}|F\rangle +\beta \sum_{y}G^{y}_1(t_0)|y\rangle]
 \nonumber\\= (\alpha \gamma  e^{-i\epsilon_0 t_0} -\beta \delta^* G^{m}_1(t_0))|F\rangle
 + \alpha\delta e^{-i\epsilon_0 t_0}  |m\rangle \nonumber\\ + \beta \gamma \sum_{y'} G^{y'}_1(t_0)|y'\rangle + \beta \delta \sum_{y' \neq m} G^{y'}_1(t_0)|m,y'\rangle.
  \end{eqnarray}
  Here, after the coherent operation, we can see that two-magnon states are also generating from one magnon states, that were absent in the last section. Now, the state  is a mixture of zero, one and two magnon states.This is the simplest possible operation where three different magnon sectors are obtained. Each of these sectors will have their own dynamics, during the further evolution using the Heisenberg Hamiltonian, in the third step. The state at a later time $t$ becomes,
 \begin{eqnarray} 
 |\tilde\psi(t)\rangle = U_{t,t_0}|\tilde\psi(t_{0^+})\rangle \nonumber= (\alpha \gamma e^{-i\epsilon_0 t}-\beta \delta^* G^{m}_1(t_0)e^{-i\epsilon_0(t-t_0)})|F\rangle+\\\sum_{y}[ \alpha \delta e^{-i\epsilon_0 t_0}G^{y}_m(t-t_0)\nonumber+ \beta \gamma X^{y}_{1}(m,t,t_0)]|y\rangle + \beta \delta \sum_{y_1,y_2;y_1 < y_2}L^{y_1,y_2}_{1,m}(m,t,t_0)|y_1,y_2\rangle.\\
 \end{eqnarray}

  where we have used  $X_1^n (m,t,t_0)=K_1^n(m,t,t_0)+H_1^n(m,t,t_0)$ to simplify the expression. 
  where, the new time-dependent two-magnon sector wave function is given by,
  \begin{eqnarray} 
 L^{y_1,y_2}_{y',m}(m,t,t_0)= \sum_{y'' > m}G^{y''}_{y'}(t_0)G^{y_1,y_2}_{m,y''}(t-t_0)+ \sum_{y'' < m}G^{y''}_{y'}(t_0)G^{y_1,y_2}_{y'',m}(t-t_0),\nonumber\\
   \end{eqnarray}
   in terms of the two-particle Green's function. The new Green's function is defined in terms of the two-magnon eigenfunctions $\psi_{p_1,p_2} (x_1,x_2)$, shown in Eq.5, and the eigenvalues $\epsilon_2 (p_1,p_2)$,  we have,
   \begin{equation} 
 G^{x'_1,x'_2}_{x_1,x_2}(t)=\sum_{p_1,p_2}\psi_{p_1,p_2}^{x_1,x_2} \psi_{p_1,p_2}^{x'_1,x'_2 *} e^{-it\epsilon_2 (p_1,p_2)}.
  \end{equation} 
  We have given the details of calculation of this Green's function both for two-magnon bound and scattering states in Appendix A.
  
  Now, the matrix elements of the RDM of the $l^{th}$ site,  as given in Equation(12) ,   are modified with
  $ \tilde{x}_l(t) = \frac{1}{2} \langle \tilde\Phi(t)|\mathbf{1}-\sigma^z_l| \tilde\Phi(t)\rangle$ and $ \tilde{y}_l(t)  =  \langle \tilde\Phi(t)|\sigma^+_l| \tilde\Phi(t)\rangle$  of the RDM can be calculated from Equation(28) and are given by, 
  \begin{eqnarray} 
\tilde{x}_l (t)  =  |\alpha \delta e^{-i\epsilon_0 t_0}G^{l}_m(t-t_0)+ \beta \gamma X^{l}_{1}(m,t,t_0)|^2+ |\beta|^2 |\delta|^2 \sum_{y'\neq l}|L^{l,y'}_{1,m}(m,t,t_0)|^2, \nonumber\\
\end{eqnarray}
and the off-diagonal matrix element is given by
\begin{eqnarray}
\tilde{y}_l (t) = e^{-i\epsilon_0 (t-t_0)}(\alpha \gamma -\beta \delta^* G^{m}_1(t_0)e^{i\epsilon_0 t_0})[ \alpha^* \delta^* G^{*l}_m(t-t_0) +\beta^* \gamma^*e^{-i\epsilon_0t_0}X^{*l}_{1}(m,t,t_0)]\nonumber\\+\sum_{y \neq l} (\alpha \beta^* |\delta|^2 e^{-i \epsilon_0 t_0}G^{y}_m(t-t_0)+|\beta|^2 \gamma \delta^*  X^{y}_{1}(m,t,t_0))L^{*l,y}_{1,m}(m,t,t_0).
 \end{eqnarray}
 After averaging over all pure states on the Bloch sphere average state transfer fidelity for $l^{th}$ site is given by,
  
\begin{eqnarray} 
\tilde{\mathscr{F}}_{l,m}(t;t_0) = \frac{1}{2} + \frac{|\gamma|^2}{6} [ |X^{l}_{1}(m,t,t_0)|^2+ 2  Re( e^{-i\epsilon_0 t}X^{*l}_{1}(m,t,t_0) ] + \frac{|\delta |^2}{6} [  \sum_{y \neq l}|L^{l,y}_{1,m}(m,t,t_0)|^2 \nonumber \\ - |G^{l}_m(t-t_0)|^2+ 2 Re(\sum_{y \neq l}e^{-i \epsilon_0 t_0}G^{y}_m(t-t_0) L^{*l,y}_{1,m}(m,t,t_0)) ].
\end{eqnarray}
 


   Unlike the previous case discussed in Section 3 here quantum state transfer will be contributed by both the magnon bound states and scattering states sectors. We can try to analyse how much contribution to the fidelity comes from two magnon scattering states and two magnon bound states. Hence, the time dependent function $G^{x'_1,x'_2}_{x_1,x_2}(t)$  can be split into two parts $G^{x'_1,x'_2}_{(B);x_1,x_2}(t)$ and $G^{x'_1,x'_2}_{(S);x_1,x_2}(t)$ corresponding to contributions from bound states and  scattering states respectively. It is known that for a chain with $N$ sites number of two magnon bound states is of the order of $N$ where as,  number of two magnon scattering states is  $O(N^2)$. \\
  
  The dependence of the state transfer fidelity on the target site location and time has been shown as density plots in Fig.3 for the Heisenberg dynamics with a unitary QDP.
  Fig. 3(a) and  Fig. 3(b) show the contribution to the state transfer fidelity from the two-magnon bound states and the scattering states respectively. We can see that the contribution of the two-magnon scattering states is larger than the contribution coming from the two-magnon bound states. This is according to our expectation, as the number of bound states is $N$, whereas there are $O(N^2)$ number of scattering  states. The difference in the fidelities $\Delta {\mathscr{F}}$ of the two states, $\tilde{\rho}(t)$ and $\rho(t)$, is shown as a density plot in Fig.3(c) as a function of $l$ and $t$. It can be seen from here that it can take both positive and negative values, corresponding to constructive and destructive interference of the propagation  of the
  the coded information at $t=0$ and  the unitary QDP occurring at $t=t_0$. The more interesting regime is where the fidelity difference is positive and is around 0.2, implying a 20 per cent
  increase in the probability of the state transfer, coming from the implementation of the QDP at a single site.  It is important to state here that this increase occurs only for the case, the
  coded state arrives at the location of QDP, viz. site $m$ at the time $t=t_0$, to take the maximum benefit of the interference. We have shown the difference in the fidelities in Fig. 3(d) as 
  a function of the target site $l$, for various values of the interaction parameter $\Delta$ of the Hamiltonian. We can see it is fluctuating around 0.1 for all parameter values, implying a similar behaviour. The amplitude of the fluctuations is maximum for the case of $\Delta=0$, corresponding to the XY model.

   \begin{figure*}[t!]
     \begin{center}
        \subfigure[]{%
            \label{fig:first}
            \includegraphics[width=0.46\textwidth]{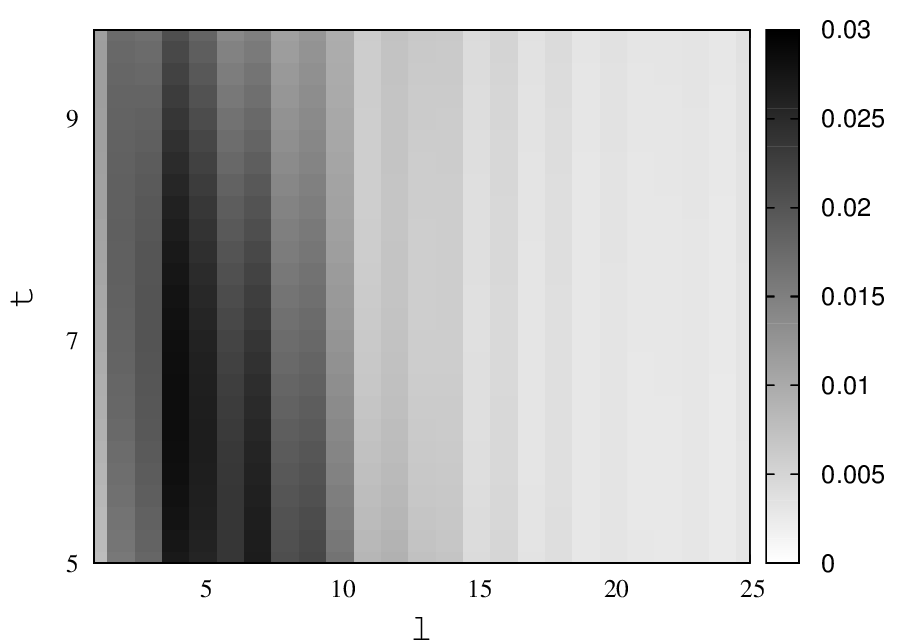}
        }%
        \subfigure[]{%
           \label{fig:second}
           \includegraphics[width=0.46\textwidth]{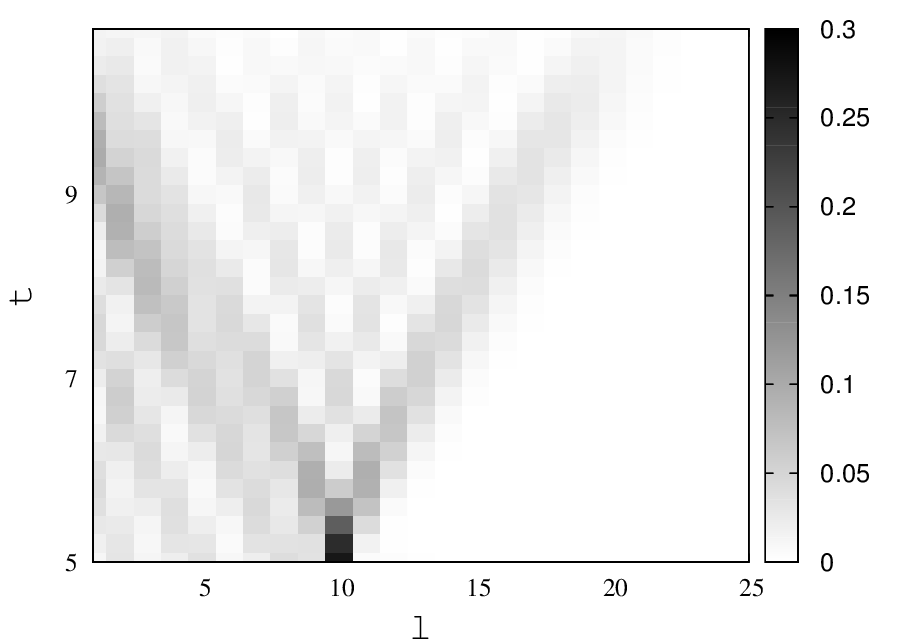}
        }\\ 
        \subfigure[]{%
            \label{fig:third}
            \includegraphics[width=0.46\textwidth]{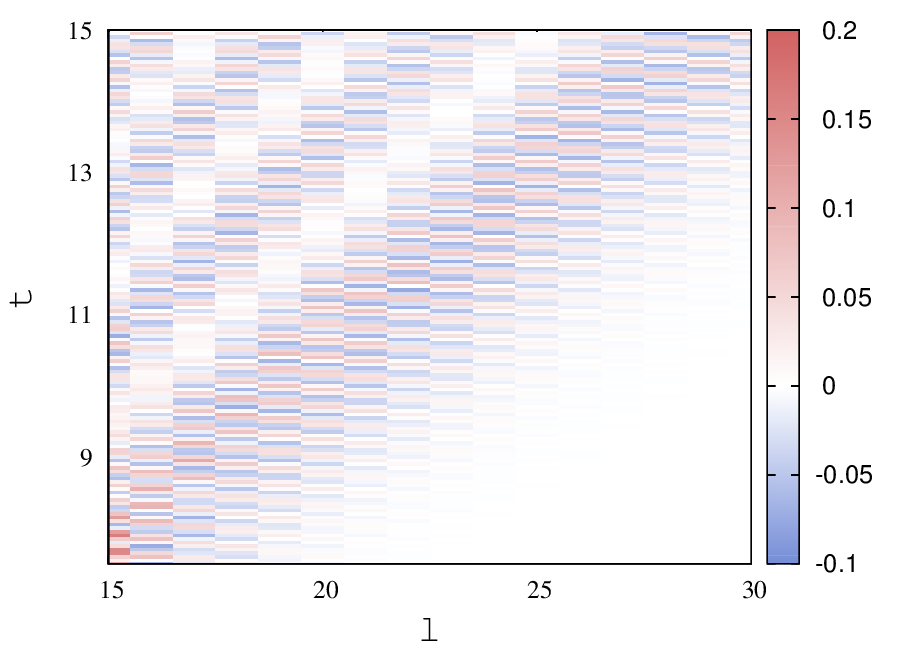}
        }%
        \subfigure[]{%
            \label{fig:fourth}
            \includegraphics[width=0.46\textwidth]{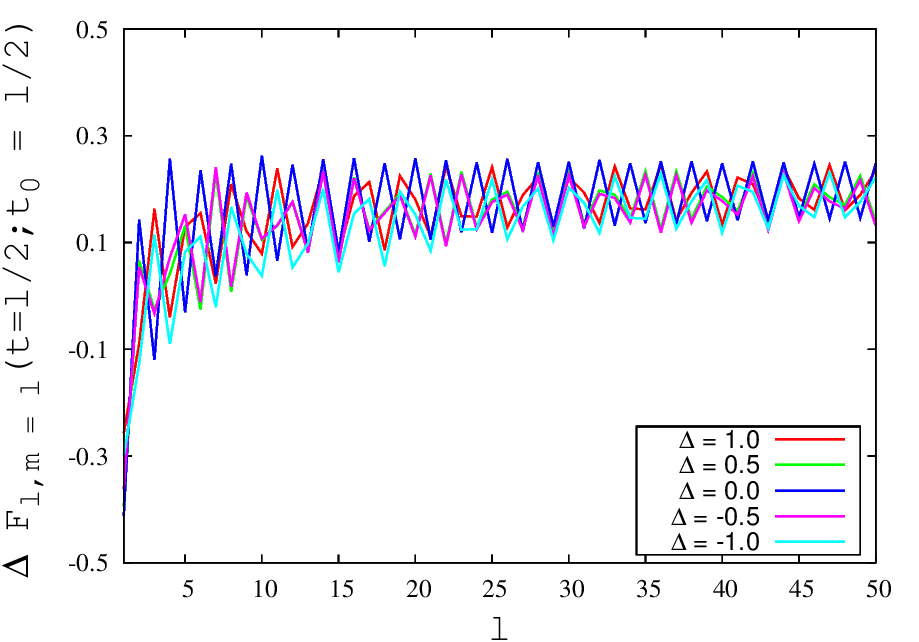}
        }%
    \end{center}
\caption{\label{fig:fig_3}{\small (a) $\frac{1}{6}\sum_{y'\neq l}|L^{l,y'}_{1,m}(m,t,t_0)|^2$ as function of site index($l$) and time ($t$) for Heisenberg dynamics with closed boundary condition with local unitary operation on the tenth site at $t_0 = 5.0$. without taking two magnon scattering states into account for anisotropy constant $\Delta = 1.0$.(b) $\frac{1}{6}\sum_{y'\neq l}|L^{l,y'}_{1,m}(m,t,t_0)|^2$ as function of site index($l$) and time ($t$) for Heisenberg dynamics with closed boundary condition with local unitary operation on the tenth site at $t_0 = 5.0$. without taking two magnon bound states into account for anisotropy constant $\Delta = 1.0$.(c) $\Delta\mathscr{F}_{l,m=15}(t; t_0= 7.5) $  as function of site index($l$) and time ($t$) for Heisenberg dynamics with closed boundary condition with local unitary operation on the tenth site at $t_0 = 7.5$  for anisotropy constant $\Delta = 1.0$. (d)$\Delta\mathscr{F}_{l,m = l}(t=l/2;t_0 = l/2)$ as function of site index $l$ for different values of anisotropy constants $\Delta$. }
     }%
   \label{fig:subfigures}
\end{figure*}

 An interesting case can be studied when the first qubit undergoes the local operation just after the evolution starts. As a consequence it can be seen from Eq. 33  that the contribution from two magnon sector becomes zero because of the term $G^y_1(0)$. The Eq. 37 and Eq. 38 then reduce to  respectively,
 
 \begin{eqnarray} 
\tilde{\mathscr{F}}_{l,m = 1}(t;t_0=0_+ )= \frac{1}{2}+ \frac{1}{6}(|\gamma|^2 -|\delta|^2)|G^{l}_1(t)|^2 +\frac{|\gamma|^2}{3} Re((e^{i\epsilon_0 t}G^{l}_{1}(t)),\nonumber\\
\tilde{\mathscr{F}}_{l,m = 1}(t;t_0=0_+ )=\frac{1}{2}+\frac{1}{6} Re((e^{i\epsilon_0t}G^{l}_{1}(t)).
 \end{eqnarray}

The last terms of the above equations are merely oscillating terms. The second equation suggests that state transfer will not take place. Setting  $|\gamma| = |\delta|$ (Hadamard gate operation)in  the first one  also shows the same.

         \begin{figure*}[t!]
     \begin{center}
        \subfigure[]{%
            \label{fig:first}
            \includegraphics[width=0.46\textwidth]{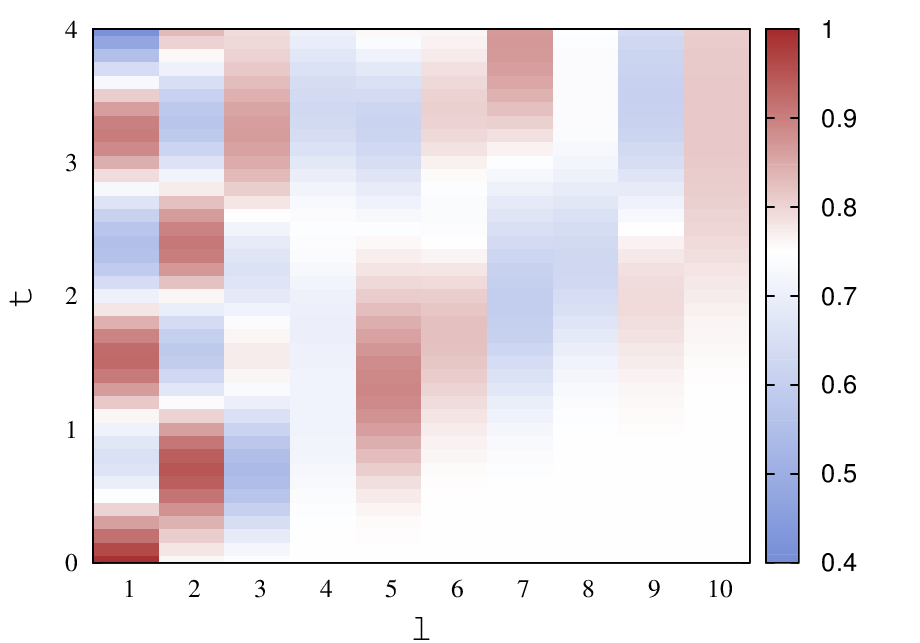}
        }%
        \subfigure[]{%
           \label{fig:second}
           \includegraphics[width=0.46\textwidth]{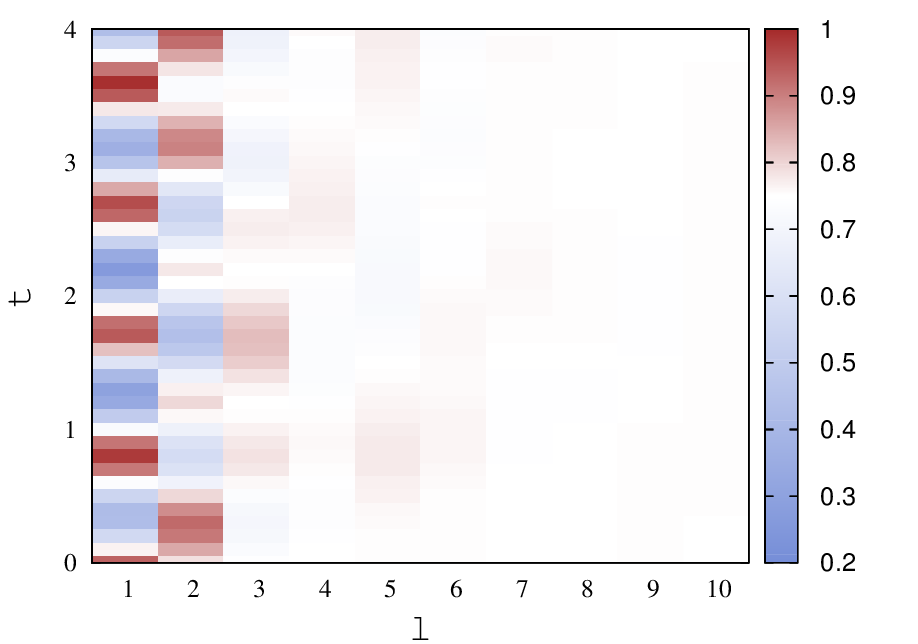}
        }\\
         \subfigure[]{%
           \label{fig:second}
           \includegraphics[width=0.46\textwidth]{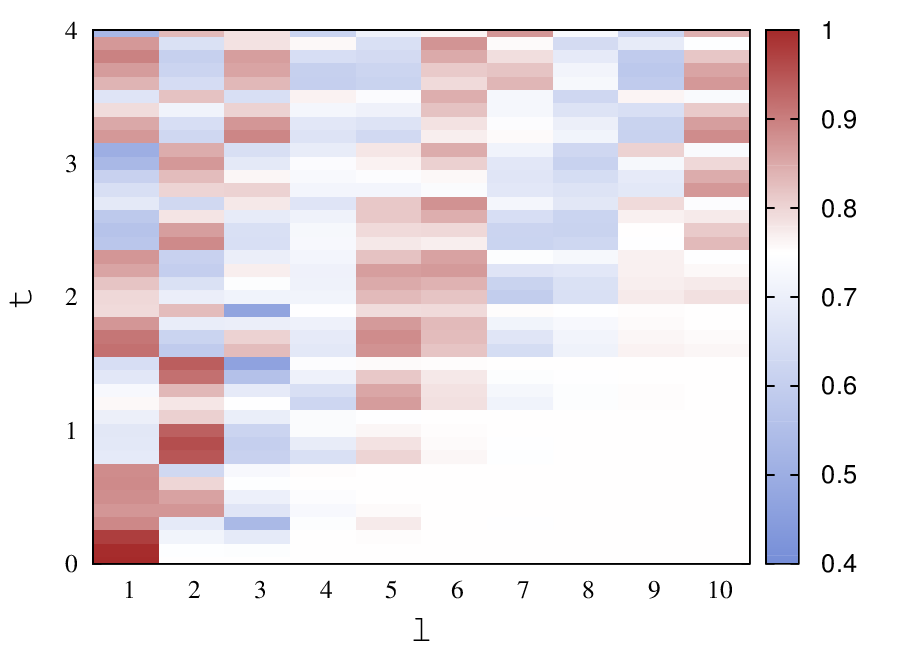}
        }%
         \subfigure[]{%
           \label{fig:second}
           \includegraphics[width=0.46\textwidth]{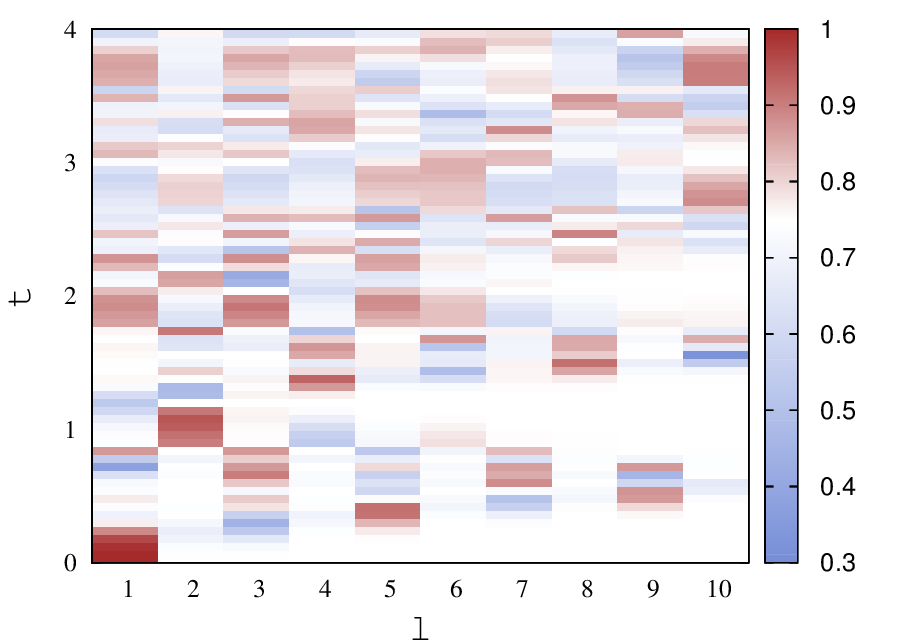}
        }%
    \end{center}
\caption{\label{fig:fig_4}{ The state transfer fidelity $\mathscr{F}_{l}(t)$ as function of site index($l$) and time ($t$) for kicked Harper dynamics with open boundary condition  with parameters \small (a)$g = 1.0$ and $\tau = 0.1$, (b)$g = 3.0$ and $\tau = 0.1$ ,   (c)$g = 1.0$ and $\tau = 0.4$, (d)$g = 1.0$ and $\tau = 0.9$ ,   $\eta = \sqrt{2}$ for each case. The initial state in all the cases is $|\psi(0)\rangle = \sqrt{0.75}|00...0\rangle+ \sqrt{0.25}|100..0\rangle$.}
     }%
   \label{fig:subfigures}
\end{figure*}

 The speed of state transfer is if the order of $J$ (here $J$ is taken to be $1/2$). Another interesting case can be studied where $t_0 = Jl = l/2$, which means that the evolution is interrupted at a particular site when the target state is passing through. In context of quantum state transfer where the target state is unknown at the receiver end one can vary the parameters $\gamma$ and $\delta$ to get the maximum fidelity. The values of $m$, $t_0$ and $t$ is set $l$, $l/2$ and $l/2$ respectively in Eq.37. Maximum value of the fidelity can be obtained when $\gamma = 0$ and $\delta = 1$ (bit flip or X gate)for large values of $l$ and the difference in fidelity with and without the local operation is given by,

   \begin{eqnarray}
\Delta\mathscr{F}_{l,m = l}(t={l\over 2};t_0 = {l\over2}) = \tilde{\mathscr{F}}_{l,m = l}(t={l\over 2};t_0  =  {l\over 2})  - \mathscr{F}_l(t={l\over 2})\nonumber \\
= \frac{1}{6} [ \sum_{y \neq l} |L^{l,y}_{1,l}(m= l, t = {l\over 2}, t_0 = {l\over 2})|^2 - |G^{l}_m(t={l\over 2})|^2 - 2 Re(e^{i \epsilon_0 l/2}G^{l}_{1}(t = {l\over2})) -1 ]. \nonumber \\
  \end{eqnarray}
  
  \section{{Local decohering process in non-integrable quantum systems}}  
In the previous sections, we have considered the effect of an intervening QDP, either non-unitary or unitary, on the background unitary Hamiltonian evolution, governed by the Heisenberg Hamiltonian, that belongs to the integrable class of dynamical systems. It will be interesting to see the effect of a QDP intervention on a unitary evolution, corresponding to non-integrable Hamiltonian systems. The signal propagation from a local QDP and its detection with a background non-integrable dynamics has been investigated in XY model with transverse and longitudinal magnetic fields\cite{ssvs}.
Moreover, there are sharp differences in the eigenvalue spacing distribution and the structure of the eigenfunctions in the two cases, that have been widely investigated.

To see the effect of QDP in non integrable systems we consider a simple model Hamiltonian with a tuneable parameter, to go continuously from completely integrable to completely non-integrable regimes. We use a one-dimensional periodically-kicked Harper model, a simple model of fermions hopping on a chain with an inhomogeneous site potential, appearing as a kick at regular intervals. The Hamiltonian is given by
 \begin{equation} 
H(t) = \sum_{j=1}^N [{\hat{c}}^{\dagger}_j { \hat{c}}_{j+1} + H.C.] +g \cos(\frac{2\pi j \eta}{N}){\hat{c}}^{\dagger}_j { \hat{c}}_{j} \sum_{n=- \infty}^{\infty} \delta({t}/\tau-n)],
  \end{equation}
  where $c_j^\dag$ is a creation operator at site $j$, $g$ is a potential strength parameter, $\eta$ is a parameter to make the potential either commensurate or incommensurate with the lattice, and $\tau$ is the kicking interval. The first term represents the kinetic energy of the fermion or hopping term, and the second term is the kicked potential energy operator. The Hilbert space for a site is two-dimensional, either occupied site or unoccupied site, thus it can be mapped to the spin language. The first term will turnout to be XY term, shown in In the Eq.2, and the second term becomes a transverse field that is inhomogeneous.  The fermion occupation can be mapped to the down spin occupation in the spin states.  Unlike the Heisenberg Hamiltonian that incorporates an interaction of down spins on neighbouring sites, there are no many-body interaction effects here. The kicked Harper model has been investigated for
  the entanglement distribution and dynamics\cite{arul3}.
The effect of the magnetic field or the potential is through  a train of kicking pulses with an interval $\tau$. For  $\tau \rightarrow 0$ the dynamics of the Harper Hamiltonian is integrable, and for large values of $\tau$ the dynamics is completely chaotic (see Figure 4 in the work of Lakshminarayan and Subrahmanyam\cite{arul3} for further details)\cite{lima}. 

 \begin{figure*}[t]
     \begin{center}
        \subfigure[]{%
            \label{fig:first}
            \includegraphics[width=0.46\textwidth]{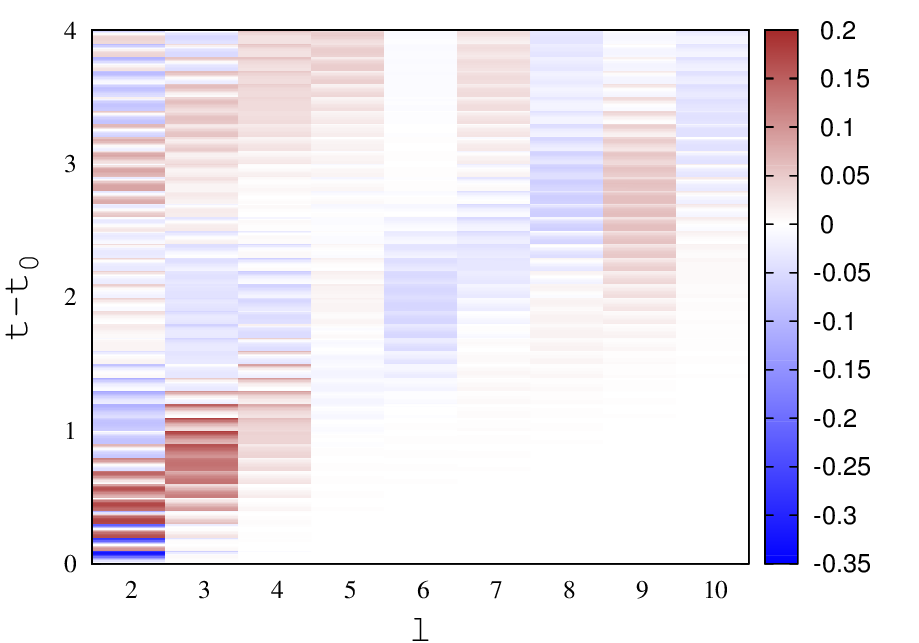}
        }%
        \subfigure[]{%
           \label{fig:second}
           \includegraphics[width=0.46\textwidth]{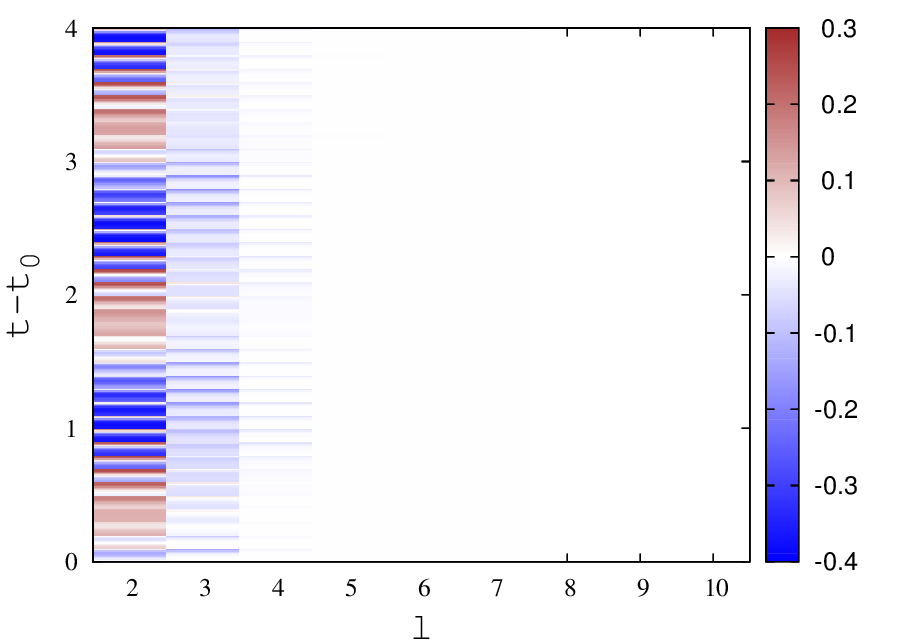}
        }\\
        \subfigure[]{%
            \label{fig:first}
            \includegraphics[width=0.46\textwidth]{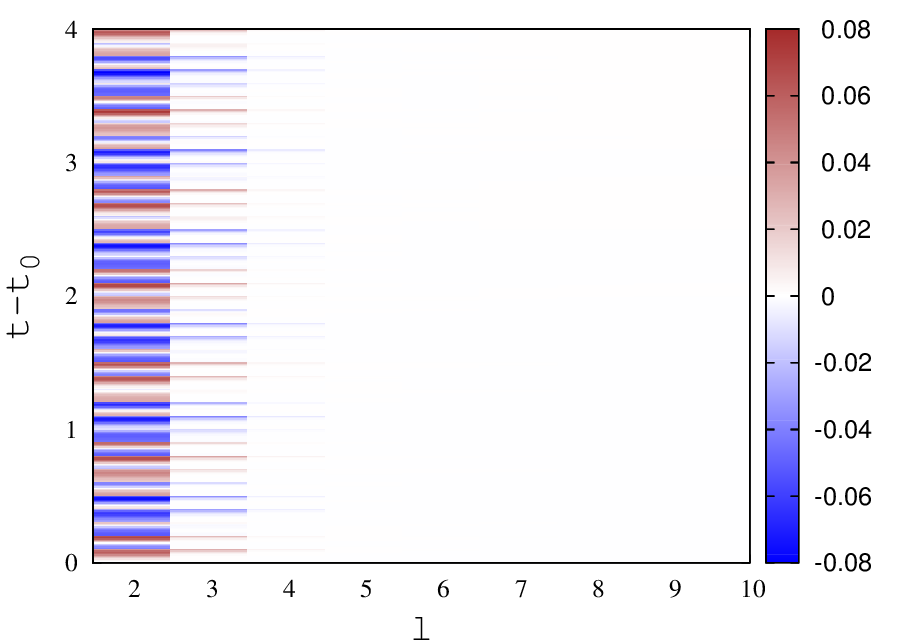}
        }%
        \subfigure[]{%
           \label{fig:second}
           \includegraphics[width=0.46\textwidth]{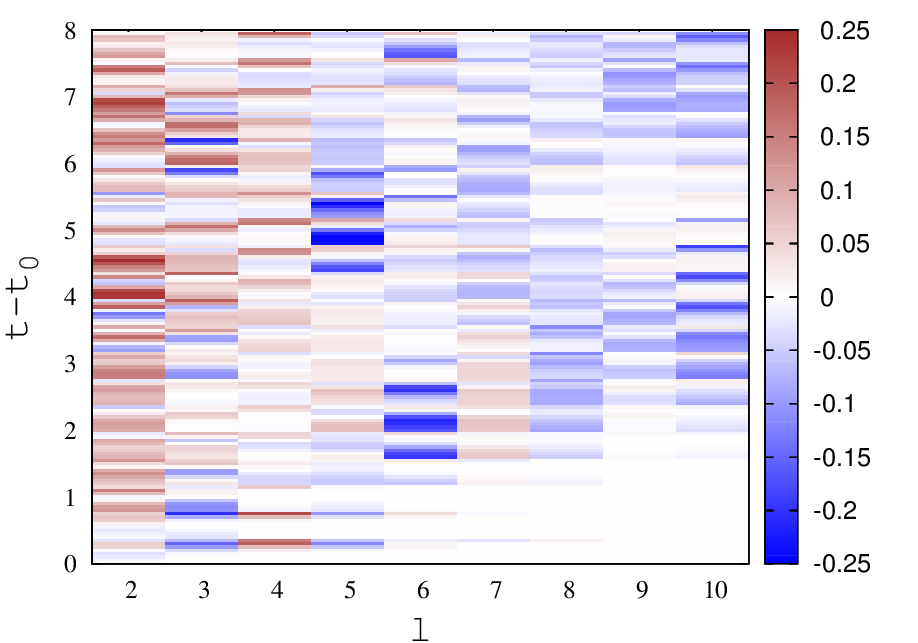}
        }\\
        \subfigure[]{%
           \label{fig:third}
           \includegraphics[width=0.46\textwidth]{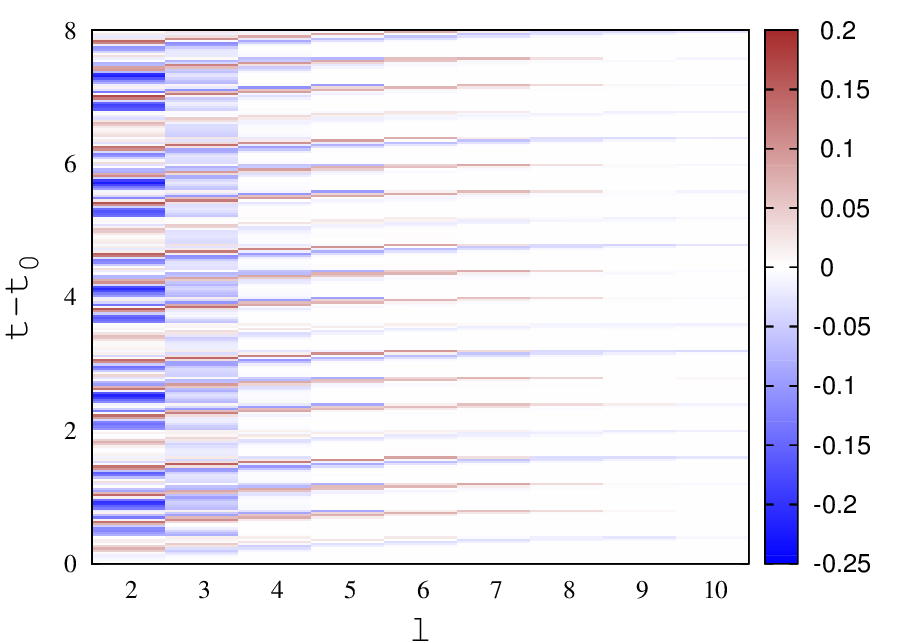}
        }%
        \subfigure[]{%
           \label{fig:fourth}
           \includegraphics[width=0.46\textwidth]{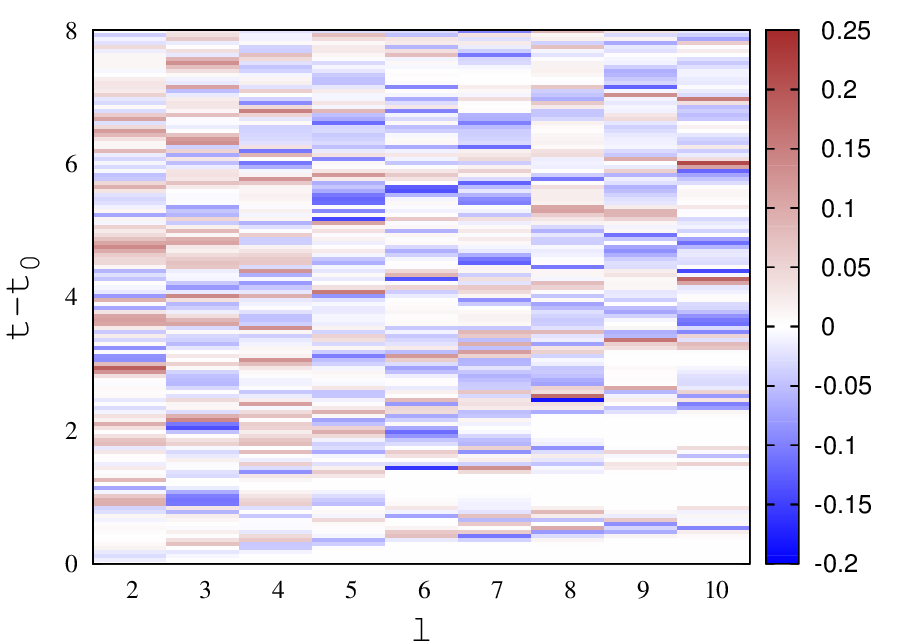}
        }%
    \end{center}
\caption{\label{fig:fig_5}{ The detector function $f_l(t)$ as function of site index($l$) and time ($t -t_0$) for kicked Harper dynamics with open boundary condition with local unitary operation on the first site after $5$ kicks ($t_0 = 5\tau$) with parameters \small (a)$g = 1.0$ and $\tau = 0.1$, (b)$g = 3.0$ and $\tau = 0.1$ , (c)$g = 10.0$ and $\tau = 0.1$, (d)$g = 1.0$ and $\tau = 0.4$, (e)$g = 2.5$ and $\tau = 0.4$, (f)$g = 1.0$ and $\tau = 0.9$.  $\eta = \sqrt{2}$ for each case.  The initial state in all the cases is $|\psi(0)\rangle = \sqrt{0.5}|00...0\rangle+ \sqrt{0.5}|100..0\rangle$.}
     }%
   \label{fig:subfigures}
   
\end{figure*}

Similar to the previous sections, we will consider an initial state $|\psi_0 \rangle = \alpha |00..0 \rangle + \beta |10..0\rangle$, a linear combination of zero-particle state and one-particle state localised at first site.  Through the time evolution, the particle can hop around to other sites. We will consider evolution at discrete  times, viz. $t=\tau^+,2\tau^+$ etc, that is at instant just after a kick. The evolution operator $U_n$ at a time just after $n$ kicks (corresponding to $t=n\tau$), is given by
 \begin{equation}
 U(n) = \left ( e^{-i \tau \sum_{j} 
c_j^\dag c_{j+1} + H.C.}  e^{-i \tau g \sum_j \cos{\frac{2\pi j \eta}{N}} {c_j^\dag c_j } }\right  )^n,
\end{equation}
where the two operator factors appearing above do not commute. This evolution operator evolves the initial state to give $|\psi_n\rangle=U(n)|\psi_0\rangle$. 

The above unitary evolution is intervened by a local instantaneous non-unitary QDP, as considered in Section 3, at time $t=n_0\tau^+$. Thus, the evolution proceeds in three steps as shown in Fig. 1. In the first step, the initial state will become the state $|\psi(n_0)\rangle$ at time $t=n_0\tau$.  In the second step, the state $\rho(n_0)$  , in analogy with
Eq. 19, will result in the state $\tilde \rho(n_0)= P_0 \rho(n_0) P_0^\dag + P_1 \rho(n_0) P_1^\dag$. The third step, the state evolves unitarily to become,
\begin{equation}
\tilde \rho(n)= U(n-n_0) (P_0  \rho(n_0) P_0^\dag  + U(n-n_0) P_1 \rho(n_0) P_1^\dag) U^\dag(n-n_0).
\end{equation}
 The quantum state transfer fidelity can also be calculated numerically, by calculating the matrix elements of the reduced density matrix  $\rho_l(n)$ for the $l$'th site, noting that $\tilde x_l$ is just
the expectation value of the fermion number operator. We have, analogous to Eq. 25,
\begin{equation}
\tilde x_l(n)= Tr ~c_l^\dag c_l ~\tilde \rho(n), ~\tilde y_l(n)=Tr ~c^\dag_l~\tilde \rho(n),
\end{equation}
where we have written the matrix elements as expectation value of operators. The state transfer fidelity can be calculated numerically using the above. The average fidelity for an initial state $|\psi_0\rangle = \sqrt{0.75}|0\rangle+\sqrt{0.25}|1\rangle$ has been shown in Fig. 4(a)-4(d) for a different values of $\tau$, $g$ as a density plot, for various target sites and times. For a smaller values of $\tau$, the dynamics is still closer to an integrable one which can be seen in Fig. 4(a). For small times, we can see larger fidelity only for small $l$. As the time increases, for larger values of $l$ get larger fidelity. This is as expected, the state is spreading at a given speed. For larger value of potential strength parameter $g$ the spreading of the state does not take place. This can be seen in Fig. 4(b) where the density plot for fidelity has been plotted for $\tau = 0.1$ and $g = 3.0$. However, for larger value of $\tau$  as shown in 4(d) even for small times too there are larger fidelities for far away sites. This implies, we cannot define a speed of propagation for the non-integrable case. This transition occurs at some intermediate value near $\tau = 0.4$ as shown in Fig. 4(c).  Similar features were seen in XY model with transverse and longitudinal fields\cite{ssvs}.

 We will define and study a detector function given by,
\begin{equation}
f_l(t=n\tau)= \tilde x_l(t)- x_l(t). 
\end{equation}
Here, the two quantities are the diagonal matrix elements of the RDM from the two states $\tilde \rho(n)$ and $\rho(n)$ evolved with and without the QDP occurrence.
This is a useful detector function to see the effect of the QDP, similar to the Loschmidt echo, but a lot simpler to calculate. This function has been studied in detail for various Hamiltonians
in conjunction with a non-unitary QDP\cite{ssvs}. One can also study similar detector functions constructed from the off-diagonal matrix elements, and the von-Neumann entropy of the
RDM. It has been see that they all show similar structure. The results for the above detector function are shown in Fig. 5(a) to Fig. 5(f) for various combination of values for $g$ and $\tau$.

For smaller values of kicking interval $\tau$ and  potential strength $g$ the detector function $f_l(t)$ shows behaviour similar to that of integrable models like Heisenberg model, XY model with transverse magnetic field \cite{ssvs}. The signal of the local QDP propagates through the chain with a finite speed which is shown in Fig. 5a). But the qualitative nature of the detector function depends on both the parameters $\tau$ and $g$. If the value of $g$ is increased the signal propagation does not take place or in other words $f_l(t)$ is zero for farther sites. This is depicted in Fig. 5(b), Fig. 5(c) and Fig. 5(e). For larger values of $\tau$  the speed of propagation abruptly becomes very high, which can taken as a signature of transition from integrability to non integrability. Fig. 5(d) shows this for $\tau = 0.4$ and $g = 1$. For $\tau = 0.9$ the signal travels very quickly to the other end of the chain and, thus, it is  difficult to define a finite speed in this case as it is shown in Fig. 6(f). Similar behaviour we have seen for the state transfer fidelity in Fig. 5. We can say that for $g>1$ and smaller value of $\tau$, the signal does not propagate to larger distances, and we cannot define a speed, as shown in Fig.5(b), (c) and (e). On the other hand, for $g=1$, and larger values of $\tau$, as seen in Fig. 5(d) and 5(f), the signal spreads out almost instantaneously to larger distances, and we cannot get a finite speed. 

\section{{Conclusions}}
 
In this paper, we have investigated the effect of a local quantum dynamical process  (QDP), both unitary and non-unitary processes,  intervening the process of quantum state transfer through a unitary evolution for both integrating and non integrating models. For Heisenberg model, that represents an integrable dynamics, we have analytically calculated 
the state transfer fidelity in terms of time-dependent Green functions. The information coded in the state of the first spin, spreads out to other location through the unitary dynamics, We have seen that the signal from the local QDP can interfere with the dynamics of the state transfer, and it changes the fidelity depending on the time and the location of the QDP. In the case of a non-unitary  QDP,  there is a small change in the state transfer fidelity, even if the QDP occurs at the location of the particular qubit where the coded state arrives at the time of QDP.  However, for a coherent or unitary QDP,  the fidelity can increase or decrease depending on the constructive or destructive interference of the propagation of the coded state, and the signal from the QDP. For appropriate location and the time, the state transfer fidelity can increase substantially.
In the case of unitary QDP,  an initial state with a combination of zero and one-magnon states, becomes a superposition of  zero, one and two magnon eigenfunctions.  The state transfer has contributions from both two-magnon bound states and scattering states, which are quantitatively calculated separately using the two-magnon Bethe ansatz wave functions.

Finally, we have investigated the dynamics of Kicked Harper model in the context of quantum state transfer. The dynamics of this model changes from integrable to non integrable by increasing the kicking interval time. We show that the spreading of the coded state is dependent on the kicking interval time. For smaller values of $\tau$ the dynamics is similar to the integrable model and larger values of $\tau$ the spreading takes place quickly which is a possible signature of non integrability.The signal propagation also depends on the potential strength parameter $g$ and $\tau$. The signal propagates with a finite speed below a certain value of $\tau$ and above that value the propagation takes place too quickly to define a speed. Whereas, for larger value of $g$ the signal does not reach the farther sites. So, the signal of the QDP gets localised.

\noindent {\bf Acknowledgement:} SS acknowledges the  financial support from CSIR, India. We thank Professor A. Lakshminarayan for useful discussion of non-integrable systems.

\vskip 1cm
\centerline{\bf Appendix A}  
\vskip 1cm

The time-dependent two-particle Green's function $G^{x'_1,x'_2}_{x_1,x_2}(t)$ has contributions from both scattering and bound states. We can  write the function as two parts, as
\begin{equation} 
 G^{x'_1,x'_2}_{x_1,x_2}(t) = G^{x'_1,x'_2}_{(B); x_1,x_2}(t) + G^{x'_1,x'_2}_{(S); x_1,x_2}(t).
  \end{equation}
  In Bethe Ansatz solution for the two-magnon eigenfunctions for a chain with periodic boundary conditions, the momenta can be parametrised by the relation,
  \begin{equation}
  p_i = 2 \cot^{-1}(2\lambda_i).
  \end{equation}
  The S-matrix in the Bethe Ansatz wave function is given in the form,
  \begin{equation}
  S(p_1,p_2) = e^{i\theta(p_1,p_2)},
  \end{equation}
  Where, the phase factor  $\theta$ is  given by,
  \begin{equation}
  \theta(p_1,p_2) =  2 \tan^{-1}\Big[ \frac{\Delta \sin[(p_1-p_2)/2]}{\cos[(p_1+p_2)/2] -\Delta \cos[(p_1-p_2)/2] }\Big].
  \end{equation}
  Bound state solutions are complex and are given by the form,
  \begin{equation}
  \lambda_1 = q + i/2 ; \lambda_2 = q - i/2.
\end{equation} 

Here, we consider the case of infinitely long isotropic ferromagnetic chain where $\Delta = 1$ to calculate the bound state contribution to state transfer fidelity. In this case the the number $q$ varies continuously within the limits $ -\infty < q < +\infty$ . The S-matrix for bound state wave function can be calculated using the solutions given above using Eq. 49,
  \begin{equation}
  e^{i\theta(\lambda_1,\lambda_2)} = e^{2i \tan^{-1}(\lambda_1-\lambda_2)}= e^{\log \frac{i-\lambda_1 + \lambda_2}{i+\lambda_1 - \lambda_2}} = 0.
  \end{equation} 
  Two magnon bound state wave function apart from a normalisation factor is given by, 
  \begin{equation}
  \psi^{x_1,x_2}_{p_1,p_2} = e^{i(p_1x_1+p_2x_2)}
  \end{equation}
  Since the momenta $p_1$ and $p_2$ are related as shown in Eq. 49,  the wave function for the bound state can be written in terms of the quantity $q$ as,
  
   \begin{eqnarray}
\psi^{x_1,x_2}_{q} = A(q) e^{2i[x_1 \cot^{-1}(2q+i)+x_2 \cot^{-1}(2q-i)]}= A(q) e^{x_1 \log (\frac{iq -1}{iq}) + x_2 \log (\frac{iq}{iq+1})}\nonumber\\= A(q) \big( \frac{q^2}{1+q^2} \big)^{(x_2-x_1)/2} e^{i(x_1+x_2) \tan^{-1}(\frac{1}{q})}.
  \end{eqnarray}
  Where, $A(q)$ is the normalisation factor.The normalisation factor for the wave function can be calculated as,
  
   \begin{equation}
 |A(q)|^{-2}
 = {\sum_{r=1}^{N-1} r (\frac{q^2}{1+q^2})^{N-r} }
   \end{equation}
    
  The function $G^{x'_1,x'_2}_{(B); x_1,x_2}(t)$ for an infinitely long ferromagnetic isotropic chain then becomes (using $x_{12}=x_2+x_2^\prime-x_1-x_1^\prime$ and $\tilde x_{12}=
  x_2-x_2^\prime+x_1-x_1^\prime$),
  \begin{eqnarray}
G^{x'_1,x'_2}_{(B); x_1,x_2}(t) =\lim_{N\to\infty}\frac{N}{2\pi}\int_{ -\infty}^{+\infty } \psi_q^{x_1,x_2}\psi_q^{*x'_2,x'_2} e^{-it \epsilon(q)} \frac{\partial p}{\partial q} dq \nonumber\\ = \frac{1}{2\pi}\int_{-\infty }^{+\infty }dq \frac{\partial p}{\partial q} \frac{1}{q^2}\Big(\frac{q^2}{1+q^2} \Big)^{x_{12}/2} e^{i\tilde x_{12}\tan^{-1}(1/q)-it\epsilon(q)},
  \end{eqnarray}
  Where, $\epsilon(q)$ is the energy associated with the state and given in the form,
   \begin{equation}
  \epsilon(q) = \epsilon_0 + \frac{2}{1+q^2}
  \end{equation}

  Two magnon scattering state wave function is given by,
   \begin{equation}
  \psi^{x_1,x_2}_{p_1,p_2} = A(p_1,p_2)e^{i\theta(p_1,p_2)/2}\big(e^{i[p_1x_1+p_2x_2-\theta(p_1,p_2)/2]}\\ -e^{i[p_1x_2+p_2x_1+\theta(p_1,p_2)/2]}\big).
   \end{equation}
  
   Where, the normalisation factor $A(p_1,p_2)$ for the wave function is given by,
  
   \begin{equation}
  |A(p_1,p_2)|^{-2} =  {4\sum_{r=1}^{N-1} r \cos^2 \big[\frac{(N-r)(p_2-p_1) - \theta(p_1,p_2) }{2}  \big] }
  \end{equation}

   The function $G^{x'_1,x'_2}_{(S); x_1,x_2}(t)$ becomes,
   
   \begin{equation}
 G^{x'_1,x'_2}_{(S); x_1,x_2}(t) =  \sum_{p_1,p_2} |A(p_1,p_2)|^2 \psi^{x_1,x_2}_{p_1,p_2} \psi^{*x'_1,x'_2}_{p_1,p_2}  e^{-i\epsilon(p_1,p_2)t} .
  \end{equation}
  
   Where,  the energy eigenvalues for Heisenberg chain with anisotropy constant $\Delta$ are given by,
       \begin{equation}
  \epsilon(p_1,p_2) = \epsilon_0 +2(\Delta -\cos p_1)+2(\Delta -\cos p_2).
  \end{equation} 
  
  For an infinitely long chain it can be shown that the magnon momenta $p_1$ and $p_2$ become independent and takes continuous values in an interval depending on the value of anisotropy constant $\Delta$. For an infinitely long chain with anisotropy constant  $\Delta = 1$ and  $\Delta$ ($-1 < \Delta < 1$) Eq. 58 becomes,
  
  \begin{equation}
 G^{x'_1,x'_2}_{(S); x_1,x_2}(t) =  \frac{1}{4 \pi^2}\int_{0}^{2 \pi}  \int_{0}^{2 \pi}   dp_1 dp_2 \psi^{x_1,x_2}_{p_1,p_2} \psi^{*x'_1,x'_2}_{p_1,p_2}  e^{-i\epsilon(p_1,p_2)t}.
  \end{equation} 
  This can rewritten as, using $\phi=\pi - \cos^{-1}(-\Delta)$, as
   \begin{equation}
 G^{x'_1,x'_2}_{(S); x_1,x_2}(t) =\frac{1}{4\phi^2}\int_{-\phi}^{\phi} dp_1 \int_{-\phi}^{\phi}   dp_2 \psi^{x_1,x_2}_{p_1,p_2} \psi^{*x'_1,x'_2}_{p_1,p_2}  e^{-i\epsilon(p_1,p_2)t}.
  \end{equation}

        \end{document}